\documentclass[journal = jctcce, manuscript = article]{achemso}
\setkeys{acs}{usetitle = true}
\usepackage{times}
\usepackage{xcolor}
\usepackage[normalem]{ulem}
\usepackage{mathptmx}
\usepackage{amsmath}
\usepackage{amssymb}
\usepackage[version=4]{mhchem}
\usepackage{lineno}
\usepackage{multirow}
\usepackage{dcolumn}
\usepackage{wrapfig}
\usepackage{cuted}
\usepackage{xr}
\usepackage{threeparttable}
\usepackage{booktabs}
\makeatletter
\newcommand*{\addFileDependency}[1]{
  \typeout{(#1)}
  \@addtofilelist{#1}
  \IfFileExists{#1}{}{\typeout{No file #1.}}
}
\makeatother

\newcommand*{\myexternaldocument}[1]{%
    \externaldocument{#1}%
    \addFileDependency{#1.tex}%
    \addFileDependency{#1.aux}%
}
\myexternaldocument{supplemental}

\newcommand{\angstrom}{\mbox{\normalfont\AA}}
\let\textDJ\DJ
\let\DJ\relax

\DeclareRobustCommand{\DJ}{%
  \ifmmode
    \mathDJ
  \else
    \textDJ
  \fi
}
\makeatletter
\newcommand{\mathDJ}{\text{\raisebox{0.25ex}{-}\kern-0.4em$\m@th D$}}
\makeatother

\title{A Reverse Non-Equilibrium Molecular Dynamics (RNEMD) Algorithm for Coupled Mass and Heat Transport in Mixtures}
\author{Cody R. Drisko}
\author{J. Daniel Gezelter}
\email{gezelter@nd.edu}
\affiliation[University of Notre Dame]{251 Nieuwland Science Hall, Department of Chemistry and Biochemistry, \\
University of Notre Dame, Notre Dame, Indiana 46556}

\begin{document}

\begin{tocentry}
\center\includegraphics[width=\linewidth]{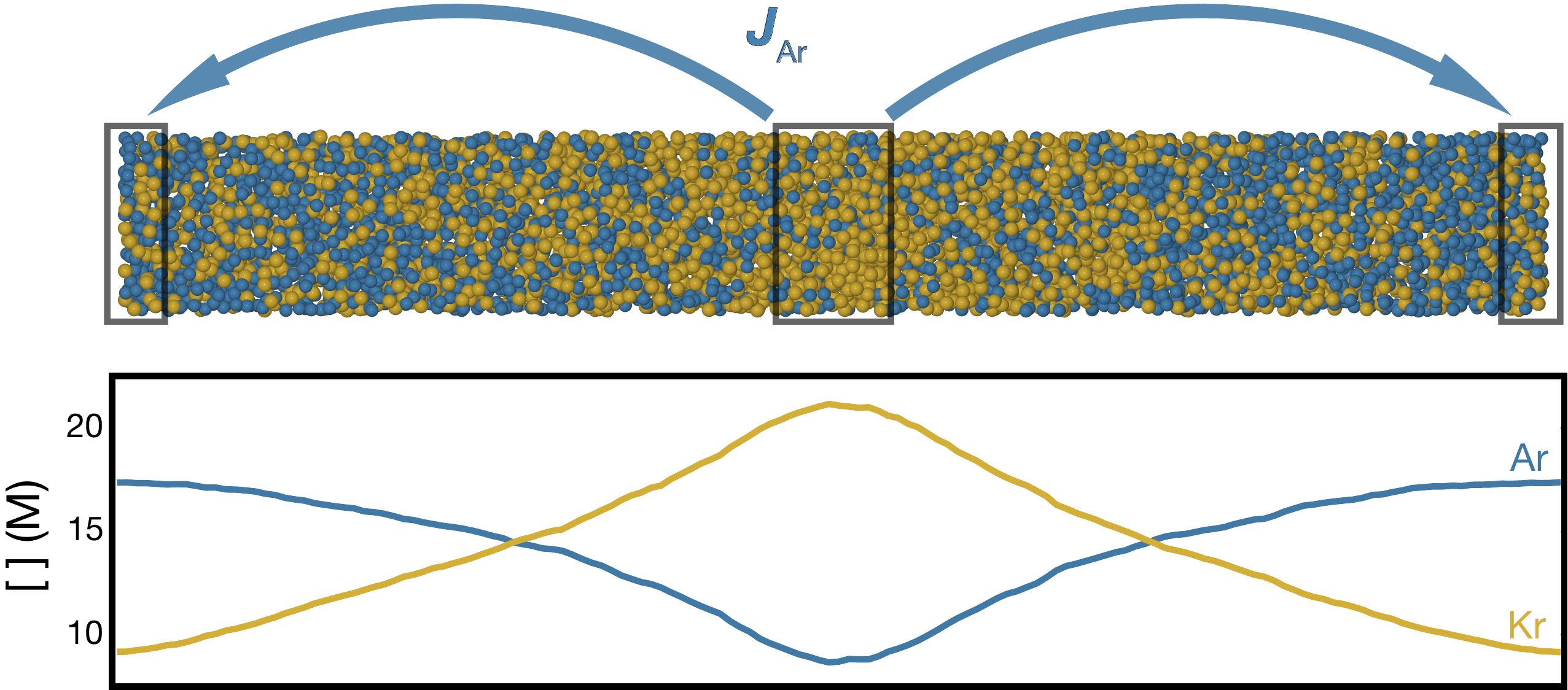} 
\end{tocentry}

\begin{abstract}
  We present a new method for introducing stable non-equilibrium
  concentration gradients in molecular dynamics simulations
  of mixtures. This method extends earlier Reverse
  Non-Equilibrium Molecular Dynamics (RNEMD) methods which use
  kinetic energy scaling moves to create temperature or velocity gradients. In the new scaled particle flux (SPF-RNEMD) algorithm, energies and forces are computed simultaneously for a molecule existing in two non-adjacent regions of a simulation box, and the system evolves under a linear combination of these interactions.  A continuously increasing particle scaling variable is responsible for migration of the molecule between the regions  as the simulation progresses, allowing for simulations under an applied particle flux. To test the method, we investigate diffusivity in mixtures of identical, but distinguishable particles, and in a simple mixture of multiple Lennard-Jones particles. The resulting concentration gradients provide Fick diffusion constants for mixtures. We also discuss using the new method to obtain coupled transport properties using simultaneous particle \textit{and} thermal fluxes to compute the temperature dependence of the diffusion coefficient and activation energies for diffusion from a single simulation. \textcolor{black}{Lastly, we demonstrate the use of this new method in interfacial systems by computing the diffusive permeability for a molecular fluid moving through a nanoporous graphene membrane.}
\end{abstract}

\maketitle
\pagebreak
\section{Introduction}
Fick's law connects a particle flux ($\textbf{J}$) of one component ($i$) in an $n$-component mixture with gradients of the mole fractions,
\begin{equation}
    \mathbf{J}_i = -c_t \sum_{j=1}^{n-1} D_{ij} \nabla x_j~~~~~~~~~ i = 1, \ldots , n-1.
    \label{eq:ficks}
\end{equation}
Here $c_t$ is the total molar concentration, $x_j$ is the mole fraction of component $j$, and $D_{ij}$ are the Fick diffusivities. In a single component or binary mixture, where there is only one independent flux, this simplifies to a single expression,
\begin{equation}
\mathbf{J}_1 = -D \nabla c_1~,
    \label{eq:fickbinary}
\end{equation}
where the diffusion coefficient depends on the molecular details (and the thermodynamic state point).

The Maxwell-Stefan (MS) model provides a generalization to Fick's law (Eq. \eqref{eq:ficks}), for multicomponent mixtures, accounting for non-ideality. It begins with gradients of the chemical potential,
\begin{equation}
    \frac{x_i}{RT} \nabla_T \mu_i = \sum_{j \neq i}^n \frac{x_i\mathbf{J}_j - x_j\mathbf{J}_i}{c_t\mathDJ_{ij}}~~~~~~~~~ i = 1, \ldots , n-1. \label{eq:MSFlux}
\end{equation}
Here $\nabla_T \mu_i = \nabla \mu_i - (\partial \mu_i / \partial T) \nabla T$, becomes the gradient of the chemical potential of species $i$ when the system is under isothermal conditions.  $\mathDJ_{ij}$ represents the MS diffusion coefficient for the $i-j$ pair. Because MS diffusion coefficients are symmetric, $\mathDJ_{ij} = \mathDJ_{ji}$, and there are $n(n - 1)/2$ diffusivities needed to fully describe mass transport. This is in contrast with with the Fick formulation, where diffusion coefficients in mixtures are not generally symmetric, so $(n - 1)^2$ diffusivities are needed to describe transport. \cite{Krishna1997, Taylor1993, Liu2013}.

In molecular simulations, chemical potentials are not simple quantities to calculate, so it is convenient to write MS diffusion (Eq. \eqref{eq:MSFlux}) in terms of spatial composition gradients. The chemical potential, $\mu_i = \mu_i^\circ + RT \ln (\gamma_i x_i)$, depends on the mole fraction and the activity coefficient, $\gamma_i$, which may depend on the other components in the mixture. Chemical potential gradients can therefore be expressed in terms of spatial gradients of the composition,
\begin{align}
    \frac{x_i}{RT} \nabla_{T} \mu_i &= \sum_{k = 1}^{n - 1} \left( \delta_{ik} + x_i \left( \frac{\partial \ln\gamma_i}{\partial x_k} \right)_{T, p, \Sigma} \right) \nabla x_k \nonumber \\ 
    &= \sum_{k = 1}^{n - 1} \Gamma_{ik} \nabla x_k~.
\end{align}
Here $\delta_{ik}$ is the Kronecker delta, the derivative now holds temperature ($T$), pressure ($p$), and all other mole fractions ($\Sigma$) constant, and $\nabla x_k$ is the composition gradient for species $k$. \citeauthor{Kirkwood1951} originally connected the thermodynamic factor, $\Gamma_{ik}$, to composition fluctuations in the grand canonical ensemble.\cite{Kirkwood1951}  Using this thermodynamic factor, Eq. \eqref{eq:MSFlux} simplifies to a convenient form,\cite{Bardow2009}
\begin{equation}
    \sum_{k = 1}^{n - 1} \Gamma_{ik} \nabla x_k = \sum_{\substack{j \neq i}}^n \frac{x_i\mathbf{J}_j - x_j\mathbf{J}_i}{c_t\mathDJ_{ij}}~, \label{eq:MSComposition}
\end{equation}
where the particle fluxes are connected to the compositional gradients.

\subsection{Diffusion via Equilibrium Molecular Dynamics (EMD)}
In single component fluids, mass transport is described by Fick's law (Eq. \eqref{eq:fickbinary}) and molecular diffusivity is characterized by the self-diffusion coefficient, $D_i$. This can be computed from the long-time behavior of the mean-squared displacement using the Einstein relation,
\begin{equation}
    D_i = \lim_{t \rightarrow \infty} ~~\frac{1}{6 N_i t}~~  \left< \sum_{l=1}^{N_i} \left( \mathbf{r}_{i,l}(t) - \mathbf{r}_{i,l}(0) \right)^2 \right>,
\end{equation}
where $N_i$ is the number of particles of type $i$, and $\mathbf{r}_{i,l}(t)$ is the position of molecule $l$ (of type $i$) at time $t$, and the angle brackets average over initial times. Equivalently, the self-diffusion coefficient can be evaluated using a Green-Kubo integral of the velocity autocorrelation function.

In contrast to self-diffusion coefficients which describe motion of individual molecules, collective mass transport is described with the Maxwell-Stefan diffusion coefficients, denoted $\mathDJ_{ij}$ in a binary $i-j$ mixture. These are most easily derived using the Onsager coefficients for mass diffusion in a two component system,\cite{Liu2013,Gullbrekken:2023vn}
\begin{equation} 
    \Lambda_{ij} = \lim_{t \rightarrow \infty} \frac{1}{N}\frac{1}{6t} \left< \left( \sum_{l=1}^{N_i} \left( \mathbf{r}_{i,l}(t) - \mathbf{r}_{i,l}(0) \right) \right) \cdot \left( \sum_{m=1}^{N_j} \left( \mathbf{r}_{j,m}(t) - \mathbf{r}_{j,m}(0) \right) \right) \right>.
\end{equation}
Here, $N$ is the total number of molecules in the system, $N_i$ and $N_j$ are the numbers of molecules of the two components, and $\mathbf{r}_{i,l}(t)$ is the position of molecule $l$ (of type $i$) at time $t$.  
In binary systems, the Maxwell-Stefan diffusion coefficient can then be expressed in terms of the three Onsager coefficients, 
\begin{equation}
    \mathDJ_{ij}^\mathrm{EMD} = \frac{x_j}{x_i} \Lambda_{ii} + \frac{x_i}{x_j} \Lambda_{jj} - 2 \Lambda_{ij} \label{eq:emd}
\end{equation}
where $x_i = N_i / N$ is the mole fraction of component $i$.

In systems with weak interactions between the two species, the Darken relation approximates the Maxwell-Stefan diffusion coefficient with the two self-diffusion coefficients,\cite{Darken1948}
\begin{equation}
    \mathDJ_{ij} \approx D_\mathrm{Darken} = x_j D_{i} + x_i D_{j}. \label{eq:darken}
\end{equation}

\subsection{Driving Flux in Molecular Dynamics Simulations}
Non-equilibrium molecular dynamics (NEMD) methods are often used to impose temperature or velocity gradients on a system,\cite{Ashurst:1975eu,Evans:1982oq,Erpenbeck:1984qe,Vogelsang:1988qv,Maginn:1993kl,Hess:2002nr,Schelling:2002dp,Berthier:2002ai,Evans:2002tg,Vasquez:2004ty,Backer:2005sf,Jiang:2008hc,Picalek:2009rz} making use of linear constitutive relations to connect the resulting thermal or momentum fluxes to transport coefficients,
\begin{equation}
    \mathbf{J}_q = -\lambda \nabla T  \hspace{0.75in}  j_z(p_x) = -\eta \frac{\partial v_x}{\partial z}.
    \label{eq:fourier}
\end{equation}
Here, $\nabla T$ and $\frac{\partial v_x}{\partial z}$ are the imposed thermal and momentum gradients, and as long as the imposed gradients are relatively small, the corresponding fluxes, $\mathbf{J}_q$ and $j_z(p_x)$, have a linear relationship to the gradients. The coefficients that provide this relationship correspond to physical properties of the bulk material, either the thermal conductivity $(\lambda)$ or shear viscosity
$(\eta)$.

In contrast, reverse non-equilibrium molecular dynamics (RNEMD) methods impose an unphysical flux between different regions of the simulation box.\cite{Muller-Plathe:1997wq,Muller-Plathe:1999ao,Patel:2005zm,Shenogina:2009ix,Tenney:2010rp,Kuang:2010if,Kuang:2011ef,Kuang:2012fe,Stocker:2013cl} The system responds by developing a temperature or velocity gradient between the two regions. The gradients which develop in response to the applied flux have the same linear response relationships to the transport coefficient of interest. Since the amount of the applied flux is known exactly, and measurement of a gradient is generally less complicated, imposed-flux methods typically take shorter simulation times to obtain converged results \textcolor{black}{for applied thermal and momentum fluxes}. At interfaces, the observed gradients often exhibit near-discontinuities at the boundaries between dissimilar materials. RNEMD methods do not need many trajectories to provide information about transport properties, and they have become widely used to compute thermal and mechanical transport in both homogeneous liquids and solids~\cite{Muller-Plathe:1997wq,Muller-Plathe:1999ao,Tenney:2010rp} as well as heterogeneous interfaces.\cite{Patel:2005zm,Shenogina:2009ix,Kuang:2010if,Kuang:2011ef,Kuang:2012fe,Stocker:2013cl}

In principle, any material property for which a linear constitutive relation is known, \textit{e.g.} Eqs. \eqref{eq:fickbinary}, \eqref{eq:MSComposition} or \eqref{eq:fourier}, may be computed by driving the appropriate flux and measuring the relevant gradient. Some of these properties may not be accessible via classical molecular dynamics but could become available with novel RNEMD techniques. In what follows, we develop a RNEMD method to create a \textbf{\textit{particle flux}} to model  diffusion, specifically in mixtures. The following section outlines a scaled particle flux (SPF) methodology which minimally perturbs the total energy of the system. This allows the method to be ``bolted on'' to existing integration methods in molecular dynamics. 

\section{Methods}
As in other RNEMD methods, we begin with the simulation cell divided into regions, including a `source' region which serves as a particle reservoir, a `sink' region where the particles will be placed, and all intervening regions which are used to record the system's response to the particle flux. The algorithm described below creates a well-defined flux, $\mathbf{J}_{\textrm{applied}}$, in a targeted component of the mixture (type $i$), and as the system evolves, concentration gradients, $\nabla c_{\{i,j\}}(\mathbf{r}, t)$, develop in response. In traditional equilibrium molecular dynamics (EMD) simulations, no external work can be done on the system, and the fluxes in Eq. \eqref{eq:MSComposition} are not independent. However, non-equilibrium simulations don't have this constraint, so a flux can be applied independently to one component of the mixture. In the case of a binary solution in which the flux is applied only to species 1, Eq. \eqref{eq:MSComposition} simplifies to
\begin{equation}
   x_2 \mathbf{J}_1 = - \mathDJ \Gamma \nabla c_1 = - D \nabla c_1. \label{eq:spfDiffusion}
\end{equation}
Here, $\mathbf{J}_1$ is the externally-applied particle flux and $\mathbf{J}_2$ is zero. Because we don't generally separate ideal from non-ideal terms, we can connect
\begin{equation}
    D = \Gamma \mathDJ~, \label{eq:MS2Fick}
\end{equation}
and there will be a linear relationship between the applied flux and the concentration gradient that allows direct computation of the Fick diffusivity.

We note that $\mathbf{J}_1$, the flux in component 1, is an unphysical, external perturbation. The system responds by creating concentration gradients, and a flow of other species in the opposite direction, but this should not be counted as an applied flux.  A useful thought experiment is to consider what would happen if particles were moved based on their prevalence in the mixture, \textit{i.e.} $\mathbf{J}_1 = x_1 \mathbf{J}_\mathrm{applied}$ and $\mathbf{J}_2 = x_2 \mathbf{J}_\mathrm{applied}$. In this case, Eq. \eqref{eq:MSComposition} predicts that no gradients will develop, independent of the composition of the mixture.

\subsection{Creating a Particle Flux}
\begin{figure}[H]
    \includegraphics[width=\linewidth]{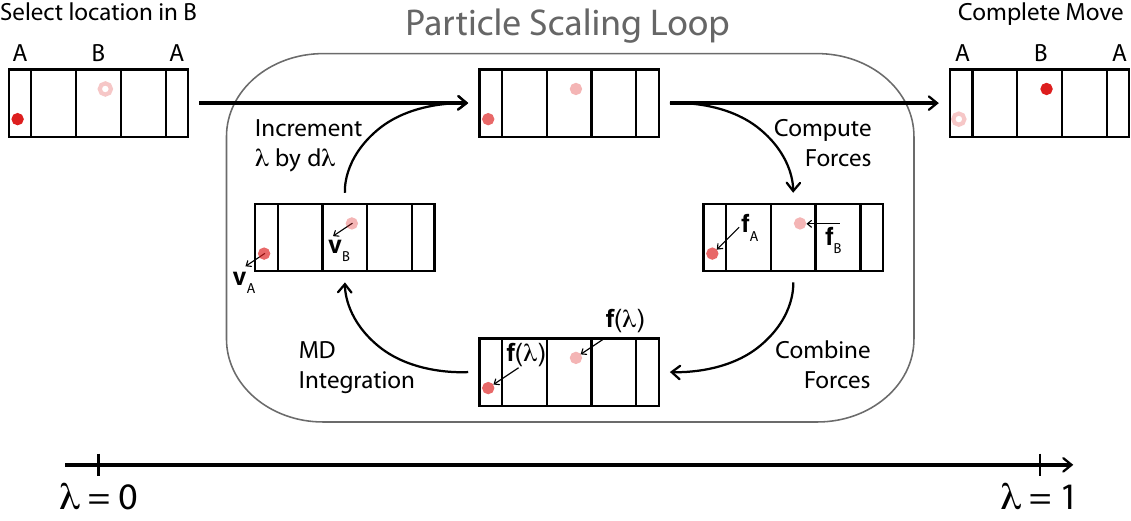}
    \caption{\label{fig:schematic} Schematic showing the main steps in the Scaled Particle Flux RNEMD method. A particle in the source bin (A) and a random location in the sink (B) are selected. In the particle scaling loop, all atomic forces using both placements of the particle are computed ($\mathbf{f}_\mathrm{A}$ and $\mathbf{f}_\mathrm{B}$). Forces are combined using the scaling function, $\mathbf{f}(\lambda) = (1-s(\lambda))~\mathbf{f}_\mathrm{A} + s(\lambda)~ \mathbf{f}_\mathrm{B}$, and the system is propagated using standard MD integration with the combined forces. After a successful step, the particle progress variable, $\lambda$ is incremented by $d\lambda$.} 
\end{figure}

To create a particle flux, a randomly selected molecule (of a specific type) is chosen to be migrated from the source region and into the sink region, determined by the directionality of the applied particle flux. This non-physical movement takes place over a sequence of time steps with a full particle exchange between source and sink occurring over an exchange period, $\tau_\mathrm{exch}$. A progress variable, $\lambda$, is introduced to represent the fraction of a particle which has been transferred between the two regions at a given time. As $\lambda$ increases from 0 (representing a particle fully present in the source region) to 1 (which represents a particle fully present in the sink region), the particle coexists in both regions, moving with forces that are determined from a linear combination of forces with the particle in both regions,
\begin{eqnarray}
    U(\mathbf{r}, \lambda) & = \left[1-s(\lambda)\right] ~ U_\mathrm{source}(\mathbf{r}) + s(\lambda)~U_\mathrm{sink}(\mathbf{r}) \\ \label{eq:scalingU}
    \mathbf{F}(\mathbf{r}, \lambda) & = \left[1-s(\lambda)\right] ~ \mathbf{F}_\mathrm{source}(\mathbf{r}) + s(\lambda)~\mathbf{F}_\mathrm{sink}(\mathbf{r})
    \label{eq:scalingF}
\end{eqnarray}
Here, $s(\lambda)$ is a function that moves smoothly from $0 \rightarrow 1$ as $\lambda$ traverses the same range. $U_\mathrm{source}$ is the potential energy with the particle fully present in the source region, and $U_\mathrm{sink}$ is the potential with it fully present in the sink region. $\mathbf{F}_\mathrm{source}$  and $\mathbf{F}_\mathrm{sink}$ are the corresponding $3N$-vectors of atomic forces. A schematic of the simulation cell showing the various steps in the method is shown in Fig. \ref{fig:schematic}.  

In practice, we are using $s(\lambda) = \lambda^k$ with $k = 3$ following a similar approach to thermodynamic integration of molecular crystals introduced by \textcolor{black}{\citeauthor{Baez:1995zt}}.\cite{Baez:1995zt}  This form of the function effectively makes very small changes to the potential when $\lambda$ is close to zero, and allows for more substantial perturbations as $\lambda \rightarrow 1$. 
Because $\lambda$ has an initial value of 0, the initial placement of a particle has no impact on the potential energy or forces. This effectively allows for a particle to be placed on top of another particle in the sink region. It is not until the particle begins to grow, \textit{i.e.} as $\lambda$ is incremented by $d\lambda$, that the system responds. The rate of change of the progress variable is directly related to the applied particle flux,
\begin{equation}
    J_{\textrm{applied}} = \frac{1}{A} \frac{d\lambda}{dt} = \frac{1}{A \tau_{exch}}
    \label{eq:appliedflux}
\end{equation}
where $dt$ is the RNEMD exchange time, $A$ is the dividing area between the two regions in the simulation cell. In rectangular simulation cells, $A = 2 L_x L_y$. 

\textcolor{black}{We note that SPF-RNEMD uses infinitesimal molecular exchange moves between two control regions, maintaining contributions from the exchanging molecule in both regions simultaneously during the duration of the exchange. A discrete approach with whole particle exchanges between control volumes was explored by \citeauthor{Heffelfinger1994} in their dual control volume grand canonical MD (DCV-GCMD) method. \cite{Heffelfinger1994} A similar scaling of Lennard-Jones contributions (using $\lambda$ to scale the repulsive wall) was proposed by \citeauthor{Shi2007} in their continuous fractional component Monte Carlo (CFC-MC) method. \cite{Shi2007}} 

\subsection{Conservation Constraints}
In other RNEMD methods for generating thermal and momentum gradients, imposed fluxes perturbed only the velocities of particles. This was usually done with additional set of constraints to conserve the total energy and linear momentum of the system. However, applying a particle flux perturbs particle positions, altering the potential energy of the system. Yang \textit{et al.}, reduced the impact of potential energy changes by using a system-wide thermostat.\cite{Yang2015} The presence of a system-wide thermostat, particularly when the system comprises materials of different heat capacities, can mask thermal gradients, as the two materials can absorb heat from the thermostat and end up at different temperatures (even if the system-wide temperature is kept constant). In the method described here, we counter the effects of potential energy changes by scaling the velocities of all particles in the two RNEMD exchange regions. In the earlier VSS-RNEMD method, velocity scaling was used to generate thermal flux between the two regions.\cite{Kuang:2012fe} Applying a particle flux jointly with a thermal flux will therefore be useful for mapping out the temperature dependence of the diffusion constant from a single simulation.

To separate the effects of particle and thermal fluxes, we introduce two scaling coefficients to the velocities of particles contained within the two RNEMD exchange regions. Total energy and the imposed kinetic energy flux act as constraints which determine the values of these coefficients. We begin by considering the net change in energy. During an increment of $d \lambda$, the net change in the combined potential energy is,
\begin{equation}
   \Delta U = \left[U_\mathrm{sink}(\mathbf{r}) - U_\mathrm{source}(\mathbf{r}) \right] \left[ s(\lambda + d\lambda) - s(\lambda) \right].
\end{equation}
To constrain total energy, the velocities of particles in the source and sink regions must be scaled to maintain an opposing change in the kinetic energy. Scaling the velocities of particles in the exchange slabs is achieved relative to the center of mass velocities of those two regions,
\begin{align}
    \textbf{v}_i &\leftarrow a \left ( \textbf{v}_i - \langle \textbf{v}_a \rangle \right ) + \langle \textbf{v}_a \rangle \\
    \textbf{v}_j &\leftarrow b \left ( \textbf{v}_j - \langle \textbf{v}_b \rangle \right ) + \langle \textbf{v}_b \rangle,
    \label{eq:velocities}
\end{align}
where $a$ and $b$ are the scaling constants for the exchange regions, $\textbf{v}_i$ and $\textbf{v}_j$ are the velocities of the unperturbed molecules, and $\langle \textbf{v}_a \rangle$ and $\langle \textbf{v}_b \rangle$ are the center-of-mass velocities for the two exchange slabs.

In previous RNEMD methods, points of intersection between constraint ellipsoids were used to determine the values of scaling coefficients to satisfy energy and momentum constraints.\cite{Kuang:2010if} Given the scaling approach above, total linear momentum in the system is trivially conserved. The scaling parameters can then be used to simultaneously impose a thermal flux $(J_q)$ and conserve total energy in the system:
\begin{align}
    a^2 &= 1 \textcolor{black}{-} \frac{\Delta U - J_q ~ A ~ \Delta t}{2 K_a - M_a \left< \mathbf{v}_a \right>^2} \label{eq:a2} \\ \nonumber \\
    b^2 &= 1 \textcolor{black}{-} \frac{\Delta U + J_q ~ A ~ \Delta t}{2 K_b - M_b \left< \mathbf{v}_b \right>^2}. \label{eq:b2}
\end{align}
Here $K_a$ and $K_b$ are the initial kinetic energies of the regions, $A (= 2L_x L_y)$ is the dividing area between the two regions, $\Delta t$ is the time interval between particle scaling moves, and $M_a$ and $M_b$ are the total masses of particles in the two regions. Note that there are effective limits on the potential energy difference of thermal flux which can be applied. Both $a^2$ and $b^2$ must be positive, and we further restrict velocity scaling during any one exchange interval to be within $0.1\%$ of the unperturbed velocities. \textcolor{black}{Readers interested in the derivations of Eqs. \eqref{eq:a2} and \eqref{eq:b2} are encouraged to consult the SI.}
 
When $J_q = 0$, both $a$ and $b$ either cool or heat their regions simultaneously.  The act of moving particles can create large potential energy gradients in both regions, placing other molecules in configurations where they experience large forces (and subsequent increases in velocity). Down-scaling velocities in both regions helps counteract this instantaneous heating. Note that is also possible to place molecules in configurations which result in lower potential energy configurations, but this is much less common, and the scaling operations would increase the local kinetic energy in those situations.
The values of $a$ and $b$, are easily calculated as the particle is shifted from source to sink in an increment of $d \lambda$. Using these values, perturbed velocities are applied to the system in a post-integration step. We note that a particle can always be placed in the sink region, but there are configurations where the change in potential energy due to particle growth will prevent further changes in $\lambda$ until velocity scaling becomes numerically feasible. In situations like this, the algorithm proceeds with the previous value of $\lambda$ until the increment results in a change to the potential energy which \textit{can} be offset by scaling the velocities in the two exchange regions. 

\section{Results and Discussion}

\subsection{Identical (but distinguishable) molecules}

As the most straightforward test case, we chose a system for which the diffusion coefficient is well known, liquid Argon represented with a Lennard-Jones potential. Lennard-Jones interaction parameters used in this study are provided in Table \ref{tab:nb}. In order to measure concentration gradients while keeping the density of the system constant throughout, individual Ar particles were tagged with either a blue or gold identifier. Thus, the applied particle flux only acts on a specific set of particles, and any property calculated from the other set measures system response. All particles are identical, but distinguishable, which leads to interesting mixing results in systems both at and near equilibrium. In these systems we simulated a range of mole fractions to test composition effects on the method and measured diffusion coefficients. 

\begin{table}[H]               
    \centering
    \caption{Nonbonded interactions used for SPF-RNEMD tests described in this section.} \label{tab:nb}  
    \bibpunct{}{}{,}{n}{,}{,}
    \begin{tabular}{ c | c  c  c   | c }      
        \toprule                 
        \text{Atom} & \text{Mass (amu)} & $\sigma$ (\AA{}) & $\epsilon$ (kcal/mol) & Source \\
        \midrule
        \ce{Ar} & 39.948 & 3.41 & 0.2381 & Ref. \protect\cite{Pas1991} \\
        \ce{Kr} & 83.80  & 3.67 & 0.3319 & Ref. \protect\cite{Pas1991} \\
        \bottomrule
    \end{tabular}
\end{table}

Five different mole fractions were used in the creation of the systems: $x_\mathrm{blue} =$ 0.1, 0.25, 0.5, 0.75, 0.9 (with $x_\mathrm{gold} = 1 - x_\mathrm{blue}$). Each of these systems started from a random configuration at a temperature of 101.8 K $(T^* = 0.85)$ and a reduced density ($\rho^* = \rho \sigma^3)$ of 0.85. This was followed by a short simulation in the canonical (NVT) ensemble to allow for structural relaxation in the original placement. Next came a pressure relaxation phase in the isothermal-isobaric (NPT) ensemble. The dimensions of the simulation boxes (and all atomic coordinates) were then scaled to the box dimensions shown in Table \ref{tab:bulk-comp}. A short thermal relaxation period followed in the canonical ensemble before a long time in the microcanonical (NVE) ensemble.

\begin{table}[H]
    \centering
        \caption{Composition of equilibrated mixtures, box dimensions, and ranges of applied particle flux values for systems used in testing the SPF-RNEMD method.  \label{tab:bulk-comp}}
    \begin{tabular}{c|c|c|ccc|c}
        \toprule
        \multirow{2}{*}{System}& \multirow{2}{*}{$N_\mathrm{molecules}$} & \multirow{2}{*}{T (K)} & \multicolumn{3}{c|}{Box dimensions}& $J_p (\times 10^{-8})$\\
	    &  & &L$_x$ (\AA\ )&L$_y$ (\AA\ )&L$_z$ (\AA\ ) & particles \AA$^{-2}$ fs$^{-1}$ \\
        \midrule
        Distinguishable Ar & 2744 & 101.8 &  &  &  & \\
        \multicolumn{1}{r|}{$x_\mathrm{blue}~=~$ 0.1, 0.25, 0.5} & & & 40.0 & 40.0 & 80.0 & 2.23 -- 6.25 \\
        \multicolumn{1}{r|}{0.75} & & & 40.0 & 40.0 & 80.0 & 2.50 -- 8.93 \\
        \multicolumn{1}{r|}{0.9} & & & 40.0 & 40.0 & 80.0 & 2.84 -- 15.6 \\
        \midrule
        Ar-Kr Mixture & 6000 & 115.77 & & & & \\
        \multicolumn{1}{r|}{$x_\mathrm{Ar}~=~$ 0.5} & & & 32.3 & 35.6 & 291.3 & 0.435 -- 3.11 \\
        \multicolumn{1}{r|}{0.75} & & & 32.1 & 35.4 & 289.8 & 0.440 -- 3.14 \\
        \multicolumn{1}{r|}{0.9} & & & 32.1 & 35.4 & 289.5 & 0.440 -- 3.14 \\
        \bottomrule
    \end{tabular}
\end{table}

\begin{figure}[H]
    \includegraphics[width=\linewidth]{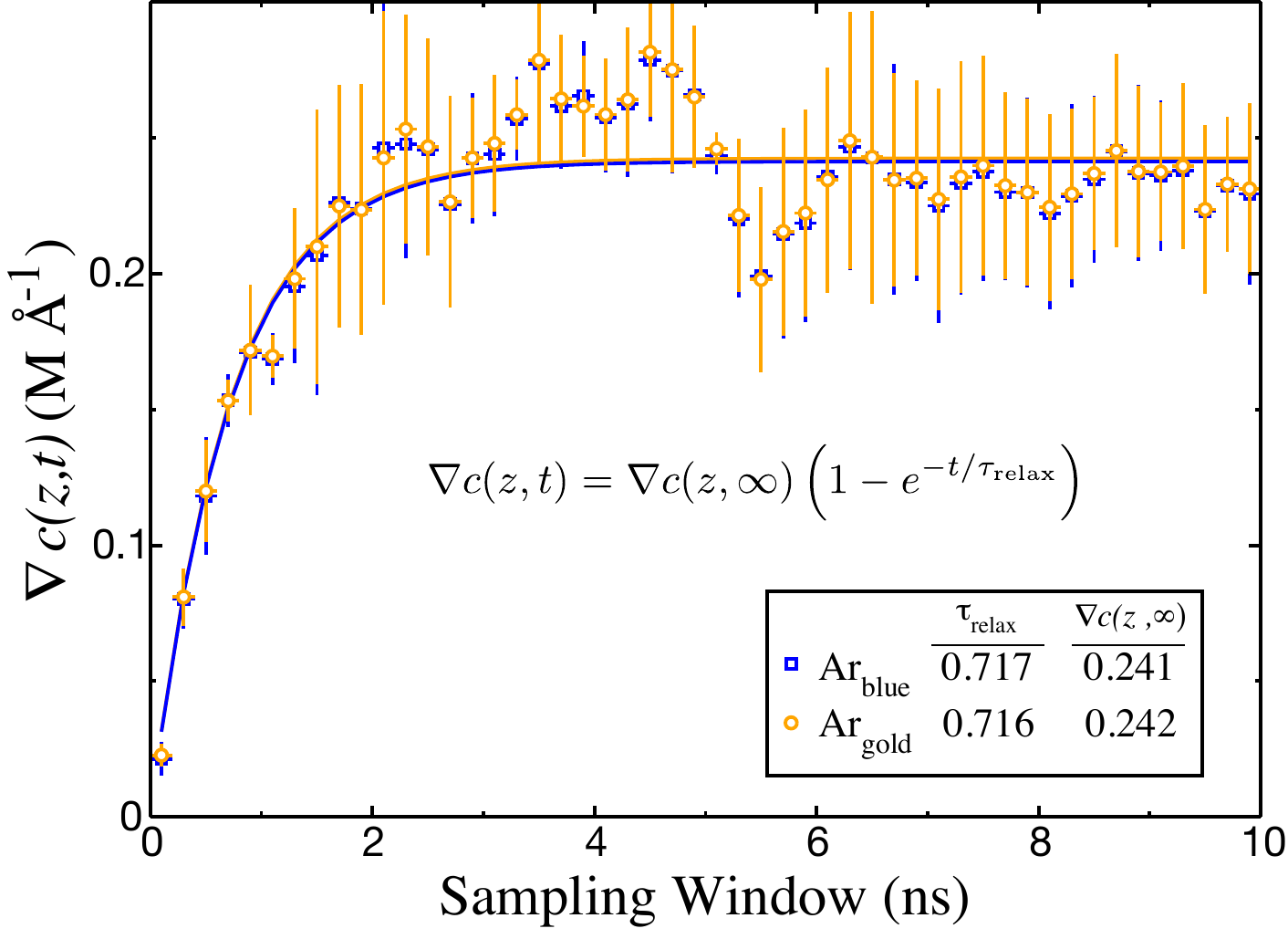}
    \caption{\label{fig:ArArStitch} For a 50:50 $\mathrm{Ar_{blue}:Ar_{gold}}$ mixture, convergence to steady-state concentration gradients is reached after approximately 4 ns with an applied flux of  $6.25  \times 10^{-8}$ $\angstrom^{-2} \mathrm{fs}^{-1}$. Data points are sampled from 200~ps windows, and 5 statistically independent replicas are used to compute average concentration gradients.}
\end{figure}

Data collection for these systems was done in the NVE ensemble for both EMD and RNEMD simulations. For the RNEMD simulations,  imposed particle fluxes between $2.23 \times 10^{-8}$ and $1.56 \times 10^{-7}$ $\angstrom^{-2} \mathrm{fs}^{-1}$  were applied to systems with a range of compositions. This flux was applied only to the molecules with the designated \textit{blue} tag. To determine when the systems had reached steady-state concentration gradients, a number of consecutive 100 ps simulations were carried out; the results of which are shown in Fig. \ref{fig:ArArStitch}. It takes approximately 4 ns under an imposed particle flux before steady state concentration gradients are reached. Therefore, these systems were simulated for a total of 6 ns, with only the last 2 ns used for data collection. To get statistically independent samples, five different configurations for each composition were equilibrated and used for data collection. The concentration gradients for the five samples were computed using linear least-squared regressions of spatial concentration profiles in the regions between RNEMD slabs. 

The thermodynamic factors $(\Gamma_{ij})$ in mixtures are connected to the correlated fluctuations in particle numbers ($N_i$ and $N_j$) within a local volume, $V$. \citeauthor{Kirkwood1951} connected these correlated fluctuations,
\begin{equation}
    G_{ij} = V \left( \frac{ \langle N_i N_j \rangle - \langle N_i \rangle \langle N_j \rangle}{\langle N_i \rangle \langle N_j \rangle} - \frac{ \delta_{ij}}{\langle N_i \rangle } \right),
\end{equation}
to integrals of the pair distribution functions,\cite{Kirkwood1951}
\begin{equation}
    G_{ij} = \int_0^{\infty} 4 \pi r^2 (g_{ij}(r) -1) dr~.
\end{equation}
The thermodynamic factor can then be expressed in terms of the Kirkwood-Buff integrals (KBIs),
\begin{equation}
    \Gamma_{ij} = \frac{x_i}{k_B T}\left( \frac{\partial \mu_i}{\partial x_j} \right) = 1 - \frac{x_i x_j \left( G_{ii} + G_{jj} - 2G_{ij} \right) } { \rho + x_i x_j \left( G_{ii}+G_{jj}-2G_{ij} \right)}~. \label{eq:thermodynamicFactor}
\end{equation}

KBIs are notoriously difficult to converge, often requiring massive simulations with significant sampling requirements. In the past few years, there have been significant advances to make these calculations more tractable under different simulation conditions and control volumes.\cite{Schnell2011,Ganguly2013,Kruger2013, Dawass2018,Dawass2019,Rosenberger2020}  In this work, we have utilized the Ganguly correction for the pair distribution function and we compute the KBIs using a cubic weighting function and integrate up to a fixed cutoff value $R$ to check for convergence,\cite{Ganguly2013}
\begin{equation}
    G_{ij}(R) = \int_{0}^{R} w(r,R) (g_{ij}(r) - 1) dr~,
\end{equation}
where
\begin{equation}
    w(r,R) = 4 \pi r^2 \left( 1 - \frac{3r}{2R} + \frac{r^3}{2R^3} \right)
\end{equation}
is an analytical weighting function for spherical control volumes.\textcolor{black}{\cite{Kruger2013}}

\begin{figure}[H]
    \includegraphics[width=\linewidth]{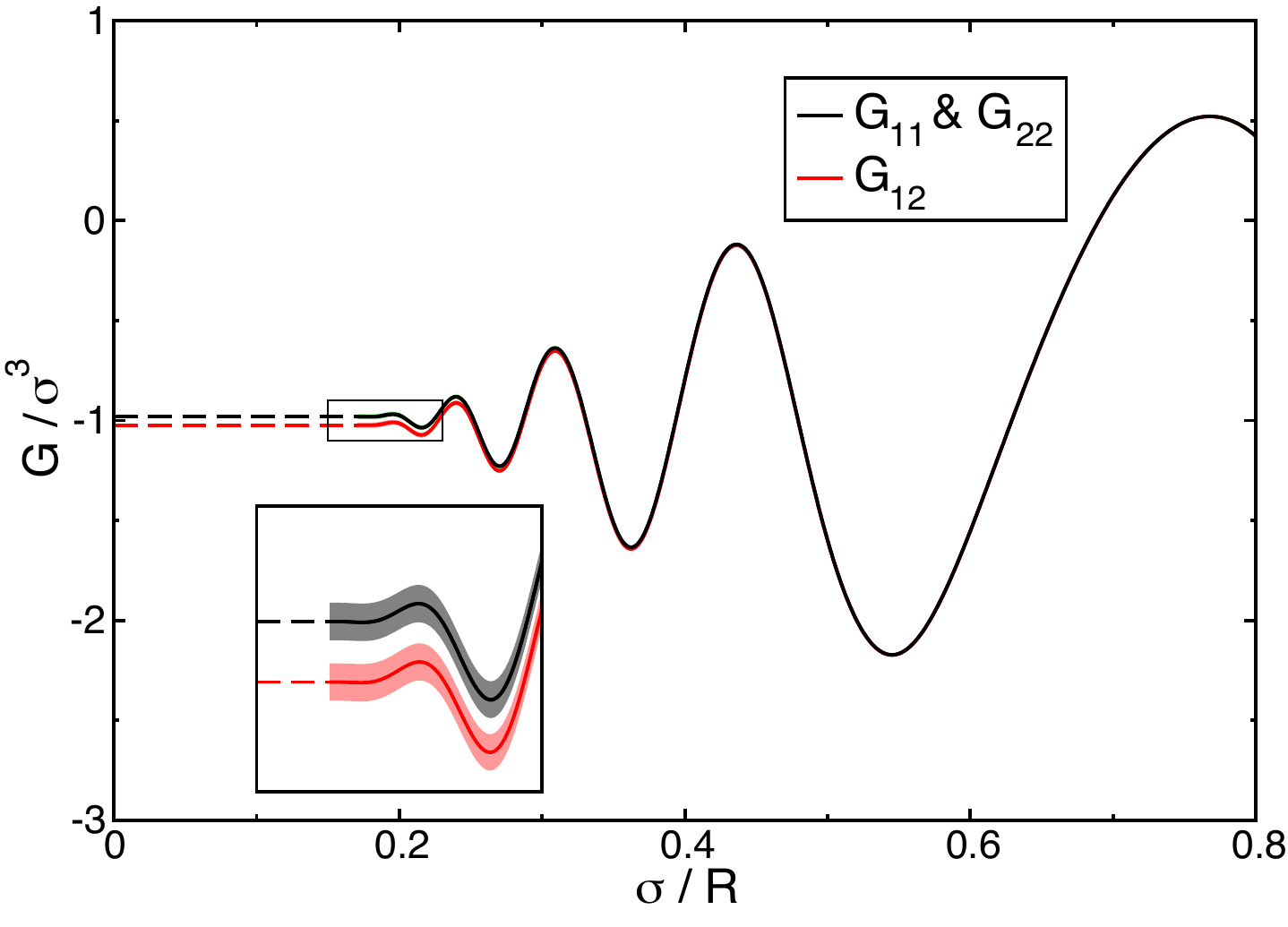}
    \caption{\label{fig:ArAr50_kb} Averaged Kirkwood-Buff integrals, $\mathrm{G}_{11}(R)$, $\mathrm{G}_{12}(R)$, and $\mathrm{G}_{22}(R)$ for the 50:50 mixture of Ar$_\textrm{blue}$ and Ar$_\textrm{gold}$.  $R$ is the outer limit of integration, and we show the convergence of as $R\rightarrow \infty$ (seen most simply as $\sigma / R \rightarrow 0$).
    The dashed lines represents the estimate of the integral in the thermodynamic limit calculated through extrapolation from the linear region. Shaded regions represent the 95\% confidence intervals for 100 statistically independent simulations. Note that for this mixture, $\mathrm{G}_{11}(R) \approx \mathrm{G}_{22}(R)$ at all distances. Differences between $\mathrm{G}_{11}(R)$ and $\mathrm{G}_{12}(R)$ indicate the contribution due to an entropy of mixing for distinguishable particles.}
\end{figure}

Once all pairwise KBIs are known, the thermodynamic factors are relatively straightforward to determine.  Sample KBIs with different spherical cutoffs for the control volume are shown in Fig. \ref{fig:ArAr50_kb}.  We used 100 independent, 1 ns simulations to compute KBIs for identical, but distinguishable Lennard-Jones Argon atoms. Values for $G_{ij}$ are extrapolated from the largest volumes to $1/R \rightarrow 0$. 

In an ideal binary mixture, the thermodynamic factor, $\Gamma_{12} = 1$, and one would expect the term $G_{11} + G_{22} - 2 G_{12}$ to vanish.\cite{BenNaim1977}  Naively, one would expect the values of $G_{11}$, $G_{22}$, and $G_{12}$ to be identical as they all measure the same pair interactions, \textit{i.e.}, Argon to Argon. 
One interesting outcome of our calculations is that for mixtures of identical, but distinguishable particles, $\Gamma_{12} \neq 1$ (see Fig. \ref{fig:ArAr50_kb}) due to an entropic driving force for mixing the distinguishable particles.
This result has an impact on the thermodynamic factor of the mixtures of distringuishable particles (Table \ref{tab:ArArDiffusionResults}). 

\begin{table}[H]
    \centering
    \caption{Diffusion constants (in $10^{-9} \mathrm{m^2 s^{-1}}$) for binary mixtures of identical (but distinguishable) Argon particles at a temperature of 101.8 K, using box geometries and particle numbers given in Table \ref{tab:bulk-comp}.\label{tab:ArArDiffusionResults}}
    \begin{threeparttable}
    \begin{tabular}{l|c|cc|ccc}
    \toprule
    & & \multicolumn{2}{c|}{$D_\mathrm{self}$}& \multicolumn{3}{c}{$D_\mathrm{Fick}$} \\ \cline{3-4} \cline{5-7}
    System & $\Gamma$ & blue & gold & Darken & EMD & RNEMD \\  
    \midrule
    \multicolumn{1}{r|}{$x_\mathrm{blue}~=~$ 0.1} & 0.92(5) & 2.2609(6) & 2.2550(4) & 2.1(1) & 2.2(1) & 2.86(5) \\
    \multicolumn{1}{r|}{0.25} & 0.97(3) & 2.2076(5) & 2.2054(4) & 2.13(7) & 2.13(7) & 2.18(3) \\
    \multicolumn{1}{r|}{0.5} & 0.98(3) & 2.2083(5) & 2.2100(5) & 2.17(4) & 2.20(7) & 2.17(6) \\
    \multicolumn{1}{r|}{0.75} & 0.97(3) & 2.2054(4) & 2.2076(5) & 2.13(7) & 2.13(7) & 2.18(6) \\
    \multicolumn{1}{r|}{0.9} & 0.92(5) & 2.2550(4) & 2.2609(6) & 2.1(1) & 2.2(1)  & 2.30(6) \\
    \bottomrule
    \end{tabular}
    \begin{tablenotes}
        \item Uncertainties in the last digits are indicated in parentheses.
    \end{tablenotes}
    \end{threeparttable}
\end{table}

\begin{figure}[H]
    \includegraphics[width=\linewidth]{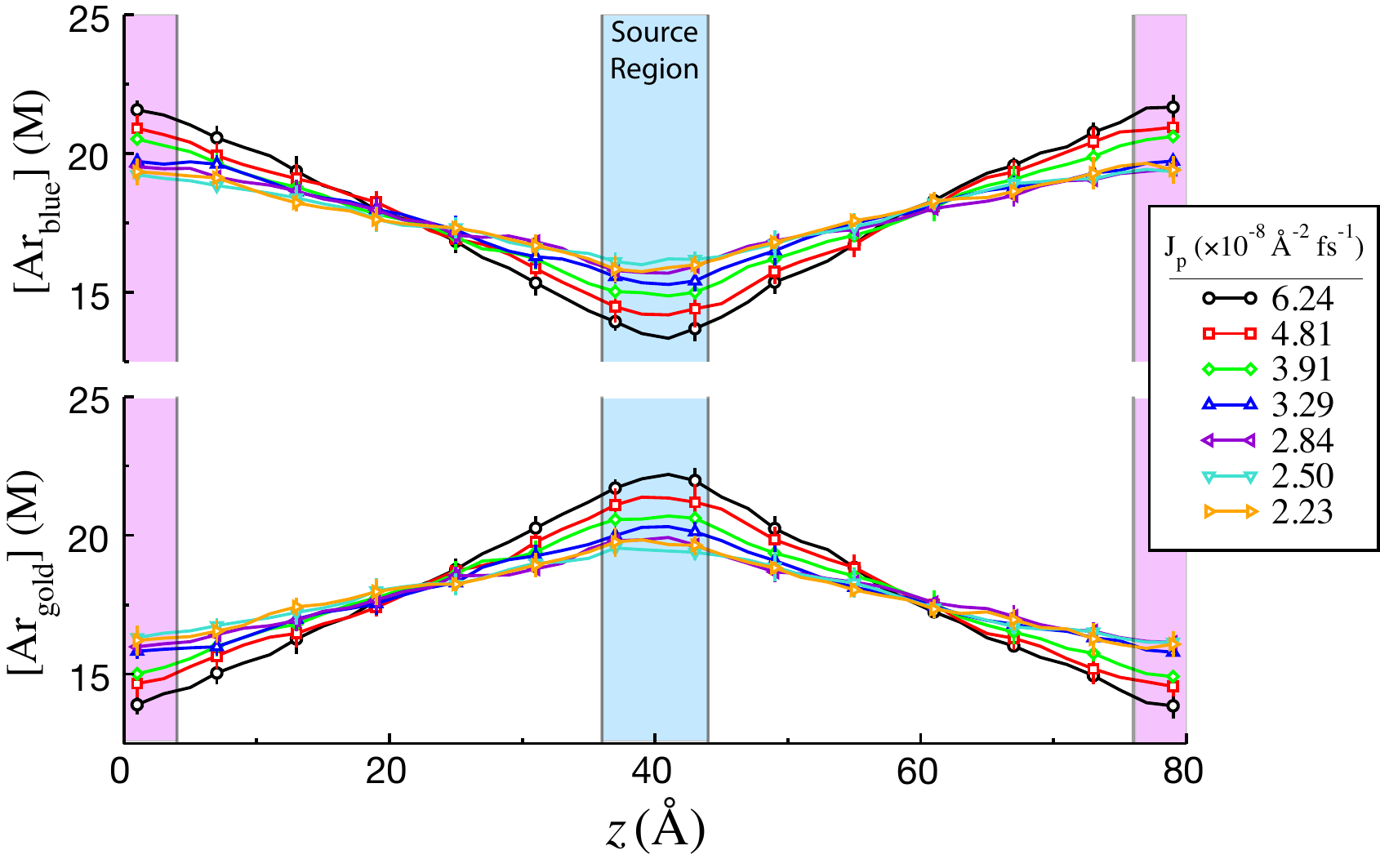}
    \caption{\label{fig:ArAr50_conc} Spatial concentration profiles for a 50:50 mixture of Ar$_\textrm{blue}$ and Ar$_\textrm{gold}$ under different applied particle fluxes ($J_p$), sampling from the last 2 ns of a 6 ns simulation. Error bars represent the 95\% confidence intervals for five statistically independent simulations.} 
\end{figure}

Fig. \ref{fig:ArAr50_conc} shows the concentration profiles of an 50:50 solution of distinguishable Argon across a range of applied particle fluxes. To compute diffusion constants, linear least squares regression was used to calculate concentration gradients from the regions between RNEMD slabs. For each applied flux, we note that two gradients (from the regions between source and sink) can be used to compute diffusion constants using Eq. \ref{eq:spfDiffusion}. These are presented in the final column of Table \ref{tab:ArArDiffusionResults}. In this table, self-diffusion coefficients were computed using mean squared displacements for particles of a given type. Fick diffusion coefficients were computed using one of three methods: 1) the Darken approximation, 2) EMD via the Onsager coefficients, or 3) using the SPF-RNEMD method (this work). Both the Darken and EMD approaches calculate the MS diffusion coefficient, so the values reported in the table are multiplied by the thermodynamic factor, $\Gamma$. \textcolor{black}{Aside from the 10:90 solution, w}e see good agreement across most mixtures between SPF-RNEMD and both the Darken approximation and the \textcolor{black}{EMD diffusivity} from the Onsager coefficients.

In the 10\% solution, at higher fluxes, the algorithm frequently fails to find a suitable particle to move from the source bin, and the concentrations in the source RNEMD region drops to zero. This  results in flux values much lower than the intended applied flux, and we have moved the system out of the linear response region. This implies that at low solute concentrations, lower applied fluxes (and longer simulation times) are going to be required to provide accurate results \textcolor{black}{that are in line with values calculated from EMD techniques}.  In the 90\% solution, the algorithm requires higher fluxes to create a measurable concentration gradient, so appropriate tuning for solute vs. solvent flux is required.  Concentration profiles for all systems (and all applied flux values) are provided in the Supporting Information (SI).

\subsection{Interdiffusion in a binary mixture of Lennard-Jones fluids}

\begin{figure}[H]
    \includegraphics[width=\linewidth]{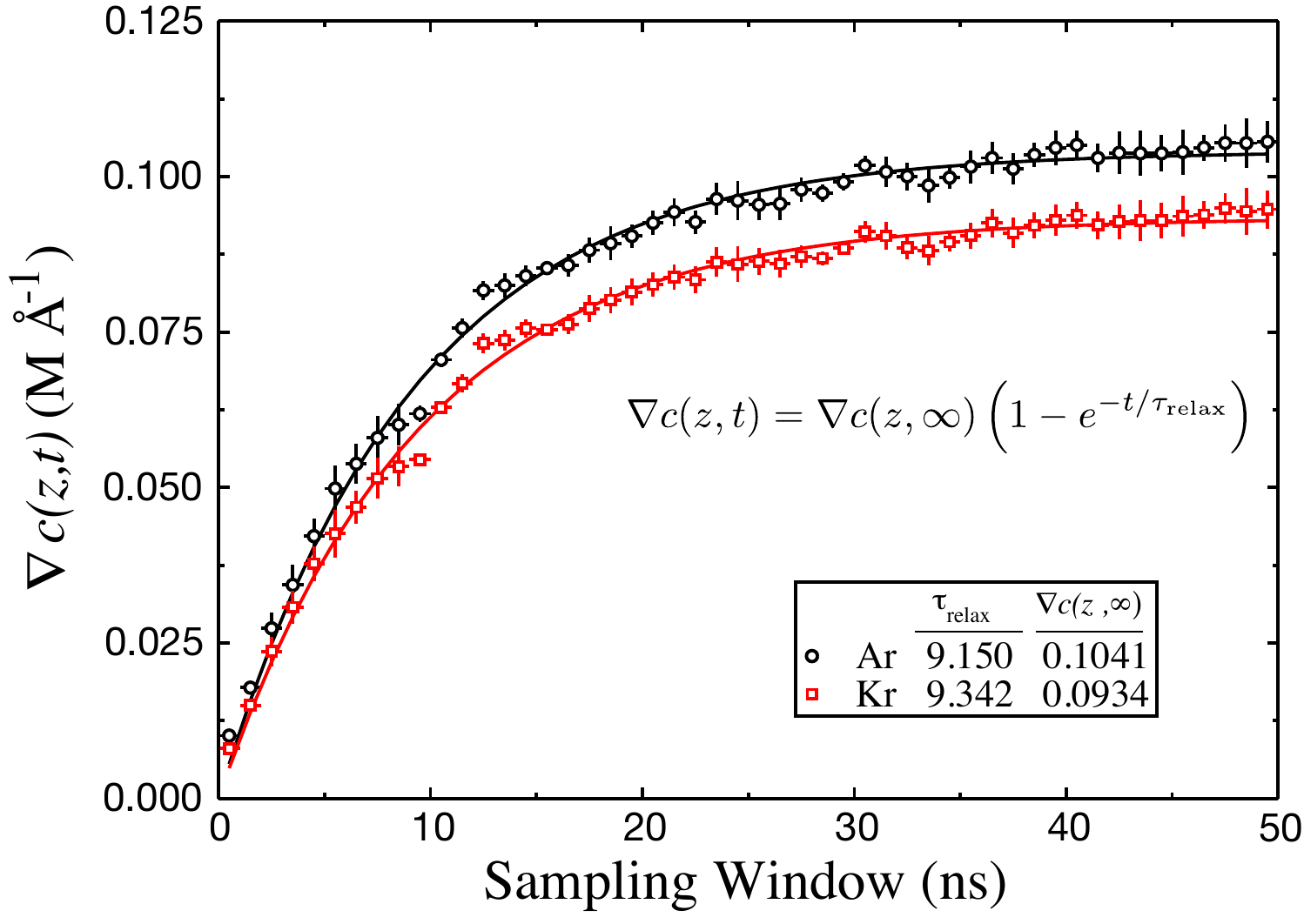}
    \caption{\label{fig:ArKrStitch} For a 50:50 Ar:Kr mixture, convergence to steady-state concentration gradients is reached after approximately \textcolor{black}{40} ns with an applied flux of  $3.11  \times 10^{-8}$ $\angstrom^{-2} \mathrm{fs}^{-1}$. Data points are sampled from 1~ns windows, and 5 statistically independent replicas are used to compute average concentration gradients.}
\end{figure}

To test the method in mixtures of dissimilar molecules, we applied the method to a mixture of Argon and Krypton atoms, again represented with a Lennard-Jones potential (parameters provided in Table \ref{tab:nb}). Simulation parameters, including box geometry, number of particles, temperatures, and applied flux values (Table \ref{tab:bulk-comp}) were identical to the ones studied in Ref. \citenum{Yang2015} which allows direct comparison between the two RNEMD approaches. Three different mole fractions were used in the creation of the systems:  $x_\mathrm{Ar} =$ 0.9, 0.75, 0.5 (with $x_\mathrm{Kr} = 1 - x_\mathrm{Ar}$). Five statistically independent replicas of all systems were started from a random configuration, and with initial velocities sampled from a Maxwell-Boltzmann distribution at 115.7 K. Equilibration was done in the same way as the distinguishable Argon simulations, but in the RNEMD \textcolor{black}{simulations}, the particle flux was applied only to the Argon atoms. Due to the larger system size (and heavier Kr atoms), convergence to steady state takes longer in this system. \textcolor{black}{Figure \ref{fig:ArKrStitch} shows that the steady state regime starts at approximately 40 ns in to the simulation. All RNEMD simulations for 50 ns, and concentration gradients were sampled only during the last 10 ns.}

\begin{figure}[H]
    \includegraphics[width=\linewidth]{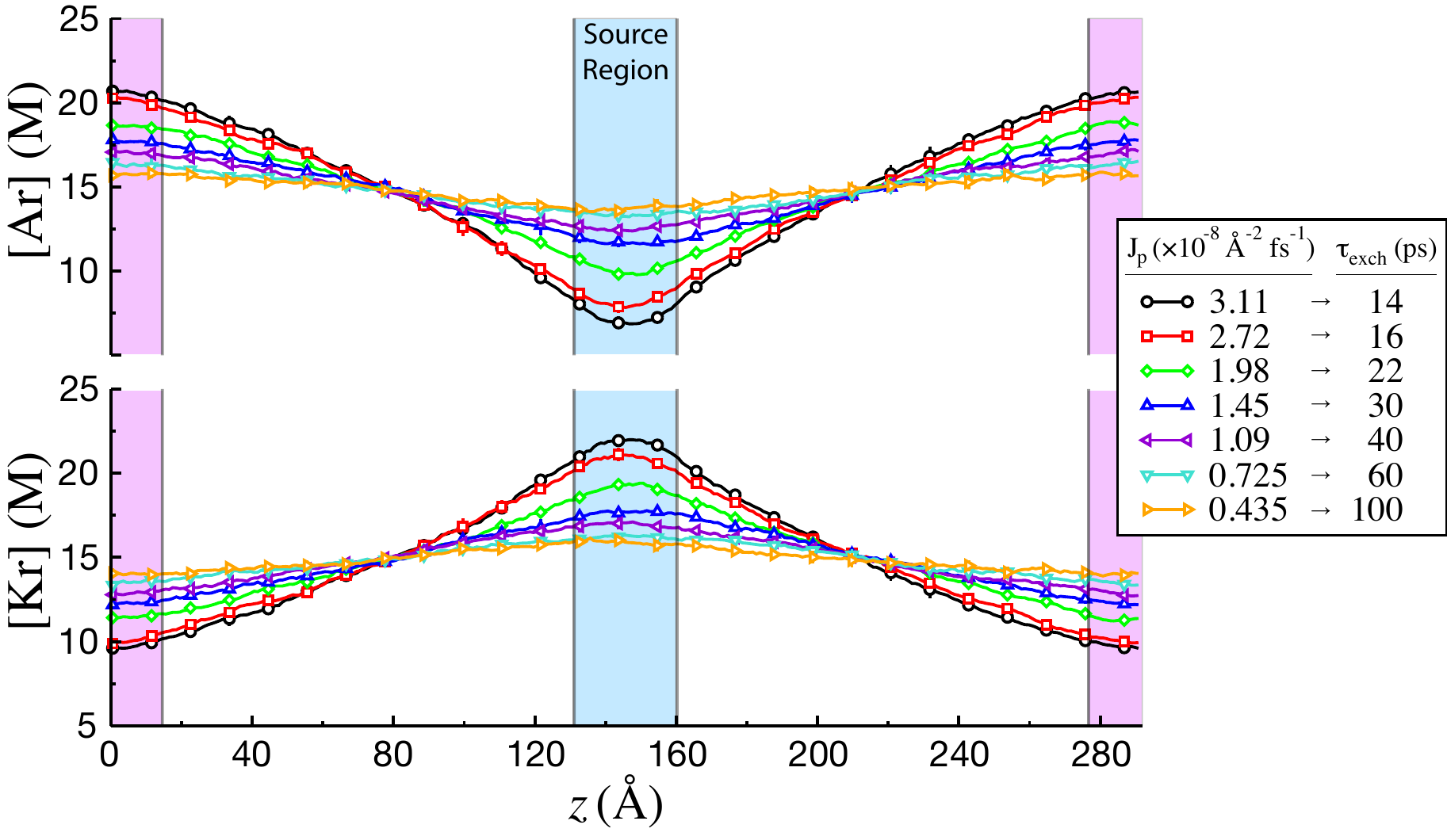}
    \caption{\label{fig:ArKr50_conc} Spatial concentration profiles for a 50:50 mixture of Ar and Kr under different applied particle fluxes ($J_p$), sampling from the last 10 ns of a \textcolor{black}{50} ns simulation. Error bars represent the 95\% confidence intervals for five statistically independent simulations, but these are generally smaller than the displayed symbols. Equivalent particle exchange periods ($\tau_\mathrm{exch}$) are also shown for each value of the applied flux.}
\end{figure}

Concentration profiles for the 50:50 mixture of Argon and Krypton are shown in Fig. \ref{fig:ArKr50_conc}. For each applied flux, two gradients (one from each of the regions between source and sink) can be used to compute diffusion constants using Eq. \ref{eq:spfDiffusion}.  These diffusion constants are presented in the final column of Table \ref{tab:ArKrDiffusionResults}. We also present self-diffusion coefficients computed using mean squared displacements for particles of a given type. Fick diffusion coefficients were computed using three separate methods: 1) the Darken approximation, 2) EMD via the Onsager coefficients, or 3) using the SPF-RNEMD method (this work). Both the Darken and EMD approaches calculate the MS diffusion coefficient, so the values reported in the table are multiplied by $\Gamma$. Concentration profiles for all other systems (and all applied flux values) are provided in the Supporting Information (SI).

\begin{table}[H]
    \centering
    \caption{Diffusion constants (in $10^{-9} \mathrm{m^2 s^{-1}}$) for binary mixtures of Argon and Krypton at a temperature of 115.7 K, using box geometries and particle numbers given in Table \ref{tab:bulk-comp}.  Fick diffusivities are compared with values calculated using a different RNEMD approach in Yang \textit{et al.}\cite{Yang2015}. \label{tab:ArKrDiffusionResults}}
    \begin{threeparttable}
    \begin{tabular}{l|c|cc|ccc|c}
    \toprule
    & & \multicolumn{2}{c|}{$D_\mathrm{self}$} & \multicolumn{4}{c}{$D_\mathrm{Fick}$} \\ \cline{3-4} \cline{5-8}
    System & $\Gamma$ & Ar & Kr & Darken & EMD & RNEMD & \textcolor{black}{Ref. \citenum{Yang2015}} \\  
    \midrule
    \multicolumn{1}{r|}{$x_\textrm{Ar}~=~$ 0.5} & 0.97(1) & 2.900(5) & 2.364(5) & 2.55(3) & 2.59(4) & 2.5\textcolor{black}{0(2)} & \textcolor{black}{2.6(2)} \\
    \multicolumn{1}{r|}{0.75} & 0.97(1) & 3.682(8) & 2.999(9) & 3.06(3) & 3.21(4) & 3.\textcolor{black}{12}(7) & \textcolor{black}{3.1(2) }\\
    \multicolumn{1}{r|}{0.9} & 0.978(7) & 4.48(1) & 3.65(1) & 3.65(3) & 3.36(3) & 3.\textcolor{black}{6(2)} & \textcolor{black}{3.8(7)} \\
    \bottomrule
    \end{tabular}
    \begin{tablenotes}
        \item Uncertainties in the last digits are indicated in parentheses. 
    \end{tablenotes}
    \end{threeparttable}
\end{table}

When compared with the identical, but distinguishable Argon mixtures, we note that this mixture shows modest disagreement between the EMD results and the other measures of Fick diffusion. 
%
The study from Yang \textit{et al.} reports diffusion coeficients of $2.63 \pm 0.19$, $3.14 \pm 0.20$, and $3.76 \pm 0.73$ (all $\times 10^{-9} \mathrm{m^2 s^{-1}}$) for the 50:50, 75:25, and 90:10 mixtures, respectively. \cite{Yang2015} Our values are all within error of these diffusion coefficients, and the trend of increasing diffusivity with increasing Argon concentration is maintained. In most cases, the differences between the EMD and RNEMD diffusion coefficients are within error estimates. We note that in contrast with previous RNEMD methods, SPF-RNEMD does not require a system-wide thermostat to maintain constant energy, which allows for heat and particle fluxes to be applied simultaneously.

\subsection{Temperature Dependence of Diffusion}
When a binary mixture is subject to thermal $(\mathbf{J}_q)$ or particle $(\mathbf{J}_1)$ fluxes, simultaneous concentration and thermal gradients may develop in the system. The phenomenological relations governing this behavior can be written in terms of the Onsager coefficients for coupled transport,\cite{NET, Rastogi:2003oq, Mortimer:1980wd, Evans:1986nx, Perronace:2002ab,Zimmermann:2022ab}
\begin{align}
\mathbf{J}_q &= -\Lambda_{qq} ~ \frac{1}{T^2}~ \nabla T - \Lambda_{q1}~ \frac{1}{T} \nabla_T (\mu_1 - \mu_2) \\
\mathbf{J}_1 &= -\Lambda_{1q}~ \frac{1}{T^2}~ \nabla T - \Lambda_{11}~ \frac{1}{T} \nabla_T (\mu_1 - \mu_2) 
\end{align}
where $\Lambda_{11}$, $\Lambda_{qq}$, $\Lambda_{1q}$, and $\Lambda_{q1}$ are the Onsager coefficients.  These relations assume that the two fluxes are equal and opposite $(\mathbf{J_2} = - \mathbf{J}_1)$.  These relations can also be written in terms of experimentally-relevant transport coefficients,\cite{Mortimer:1980wd,Evans:1986nx, Reith2000, Perronace:2002ab, Zhang2005, Zimmermann:2022ab}
\begin{align}
\mathbf{J}_q &= -\lambda \nabla T -  k_B T^2 ~ \Gamma ~ D_T^D ~\nabla c_1 \label{eq:phenomenological1} \\
\mathbf{J}_1 &= -c_t x_1 x_2 D_T^S \nabla T - D \nabla c_1
\label{eq:phenomenological2}
\end{align}
where $\lambda$ is the thermal conductivity, $D$ is the mutual (Fick) diffusion coefficient, $D_T^S$ is the thermal diffusion of the Soret type, and $D_T^D$ is the Dufour coefficient. We have also utilized the definition of the thermodynamic factor $(\Gamma)$ for a binary mixture (see Eq. \eqref{eq:thermodynamicFactor}).

If the mixture is evolving solely under a heat flux, it will reach steady state conditions and the concentration gradients will stabilize. In this case, the net flux in species 1 is zero $(\mathbf{J}_1 = 0)$.  Simultaneous measurement of both concentration and thermal gradients therefore allows for the calculation of the Soret coefficient using Eq. \eqref{eq:phenomenological2},
\begin{equation}
    s_T = \frac{D_T^S}{D} = - \frac{1}{c_t x_1 x_2} \left(\frac{\nabla c_1}{\nabla T}\right)~.
    \label{eq:soretCoefficient}
\end{equation}
        
\begin{figure}[H]
    \includegraphics[width=\linewidth]{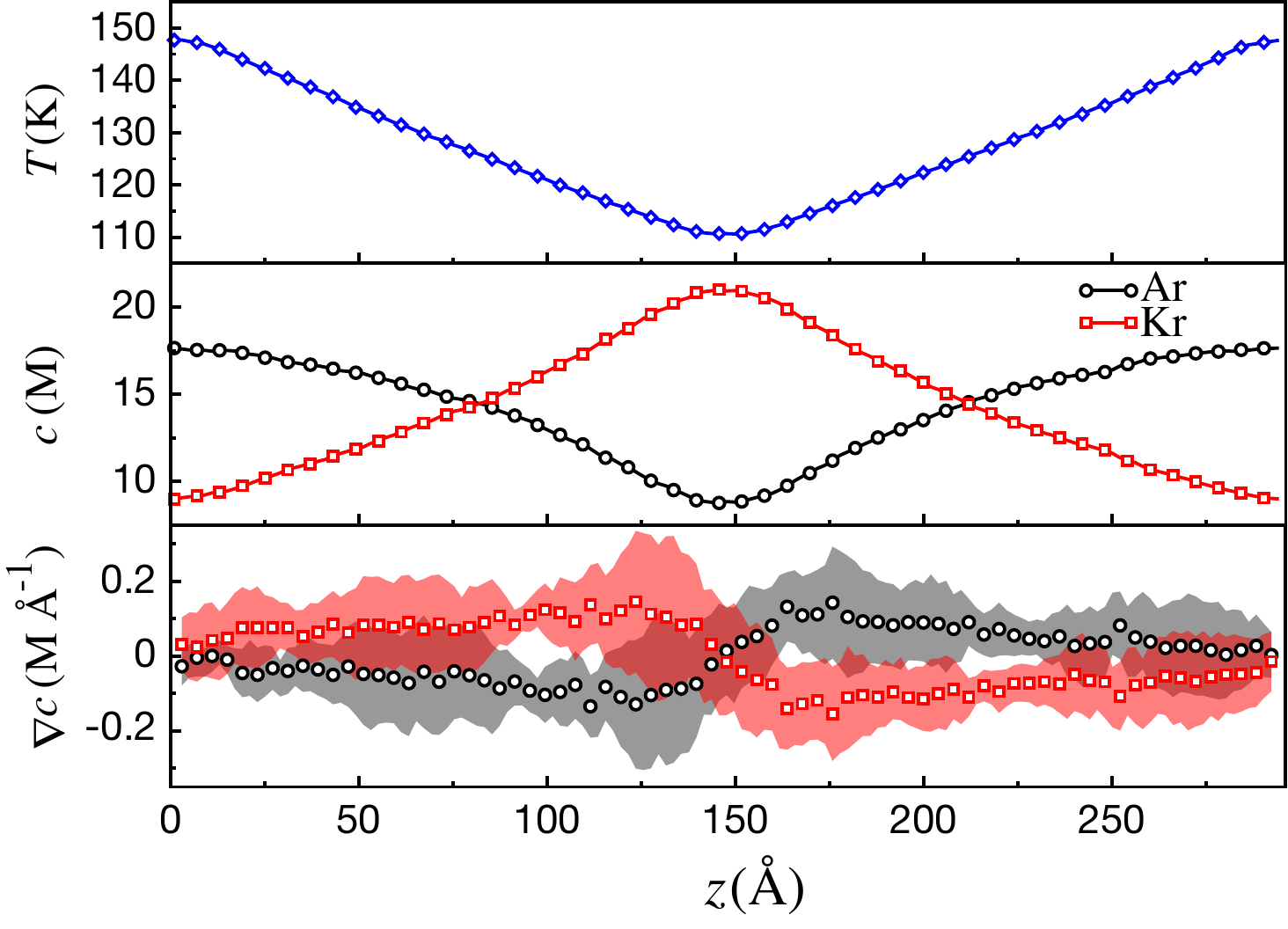}
    \caption{\label{fig:SimultaneousFlux} 
    With a combination of a thermal and a particle flux, a simulation can exhibit both a temperature (top) and a concentration (middle) gradient. The value of $\nabla c$ depends on the local temperature, so it varies along the same axis. This derivative can be computed using finite difference approximations (lower) and can be used to calculate $D$ versus T (see Fig. \ref{fig:DvsT}). } 
\end{figure}

With SPF-RNEMD, the ability to \textit{simultaneously} impose a thermal flux, $\mathbf{J}_q$, in addition to the particle flux, $\mathbf{J}_1$, means that during a single simulation, the temperature can be made to vary as a function of the box coordinates while also probing diffusive transport. The imposed thermal flux creates a linear thermal gradient as in the top panel in Fig. \ref{fig:SimultaneousFlux}.  Each point along $z$ represents a small portion of the simulation being carried out at temperature $T(z)$ with local concentration profiles and gradients.  At steady state, all of these subregions are experiencing the same particle flux through their boundaries.
The local temperature as well as the local concentration profile in the middle panel of Fig. \ref{fig:SimultaneousFlux} can then be used to map out diffusion constants over a range of temperatures using Eq. \eqref{eq:phenomenological2},
\begin{equation}
    D[T] = \frac{- x_2 ~ J_1}{c_2 x_1 x_2\ s_T \nabla T + \nabla c_1} \label{eq:activatedDiffusion}
\end{equation}
where the quantities measuring composition and temperature, $x_1$, $x_2$, $c_2$, $T$ (and their gradients) all depend on the spatial coordinate $z$.  We have also used the definition of the Soret coefficient in Eq. \eqref{eq:soretCoefficient} to simplify this expression. 

\begin{figure}[H]
    \includegraphics[width=\linewidth]{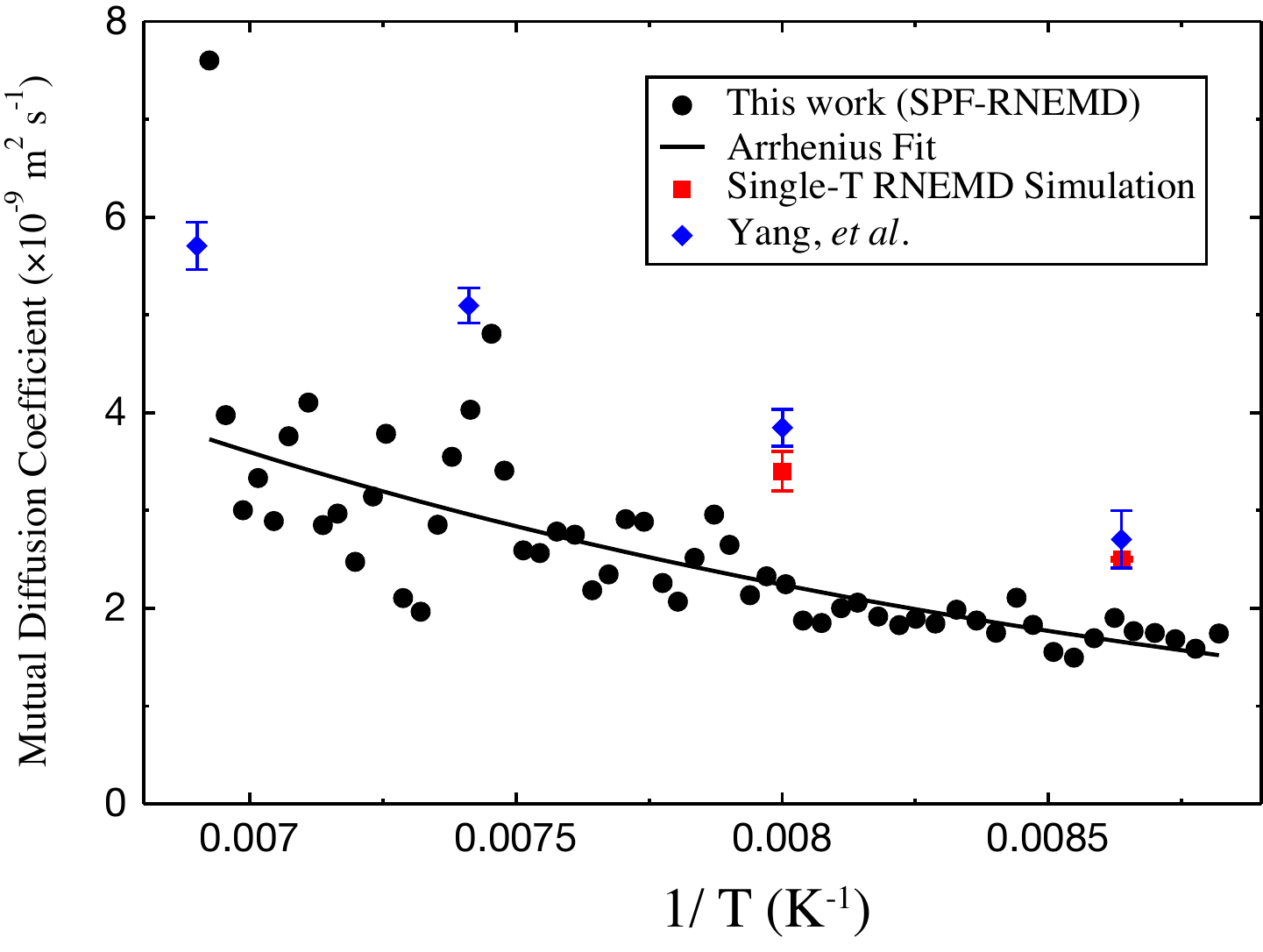}
    \caption{\label{fig:DvsT} The temperature dependence of the mutual diffusion coefficient generated from a single SPF-RNEMD simulation. Arrhenius behavior is observed for the temperature range being tested.} 
\end{figure}

To test SPF-RNEMD with simultaneous heat and particle fluxes, a set of simulations were carried out using a 6000 atom Ar/Kr mixture ($x_\mathrm{Ar} = 0.5$) at 125 K and in a simulation cell of dimensions $32.7 \times 36.1 \times 295.3~\angstrom$. Equilibration procedures were the same as in the previous sections. Similar to the other Ar/Kr systems, these simulations were run for \textcolor{black}{50} ns, with only the last 10 ns being used for data collection. The data shown in Fig. \ref{fig:SimultaneousFlux} is an average of five statistically independent simulations operating under an applied thermal flux, $J_q = 7.5 \times 10^{-7}$ kcal mol$^{-1}$ \AA$^{-2}$ fs$^{-1}$ (520 MW m$^{-2}$) and a simultaneous particle flux, $J_\mathrm{Ar} = 1.93 \times 10^{-8}$ \AA$^{-2}$ fs$^{-1}$. 95\% confidence intervals in the concentration gradients are indicated with shaded regions, but for temperature and concentration these are smaller than the displayed symbols. This data was used to generate all of the diffusion coefficients in Fig. \ref{fig:DvsT}. 

Diffusion coefficients were calculated from Eq. \eqref{eq:activatedDiffusion} with a Soret coefficient $(s_T = 3.65 \times 10^{-3}~ \textrm{K}^{-1})$. This constant was determined via a set of short RNEMD simulations to which only a heat flux was applied, and the SI contains more details on those simulations. The resulting diffusion coefficients are relatively noisy, but the temperature dependence appears to be primarily exponential vs. $1/T$, and this allows estimation of the Arrhenius activation energy for diffusion, 
\begin{equation}
    D(T) \approx D_o ~ e^{-E_a / RT}~.
\end{equation}
The activation energy obtained by fitting these diffusion coefficients, \textcolor{black}{$E_a = 3.93 \pm 0.38 \mathrm{~kJ/mol}$}, agrees well with other experiments.\cite{Kestin1984, Ghimire2017} We note that the temperature-dependent diffusion coefficients obtained from a single multi-flux simulation underestimate diffusivity when compared with those obtained from single-temperature simulations, using both our SPF-RNEMD method as well as those from Yang \textit{et al.}\cite{Yang2015}  However, the numerical calculation of concentration gradients in Fig. \ref{fig:SimultaneousFlux} is performed over a series of 2 \AA~ regions.  Narrowing the error bands in concentration gradients would require significantly more sampling (or a much larger simulation cell). In the time required for a single-temperature simulation to calculate a diffusion coefficient, application of simultaneous heat and particle flux using SPF-RNEMD generated diffusion coefficients over a wide range of temperatures.  We note that it should also be possible to use a simultaneous application of a thermal and a particle flux to map out the composition dependence of thermal conductivity, $\lambda(x_1)$.

\color{black}
\subsection{Molecular Diffusion Across A Semi-permeable Membranes}
\begin{figure}
    \includegraphics[width=\linewidth]{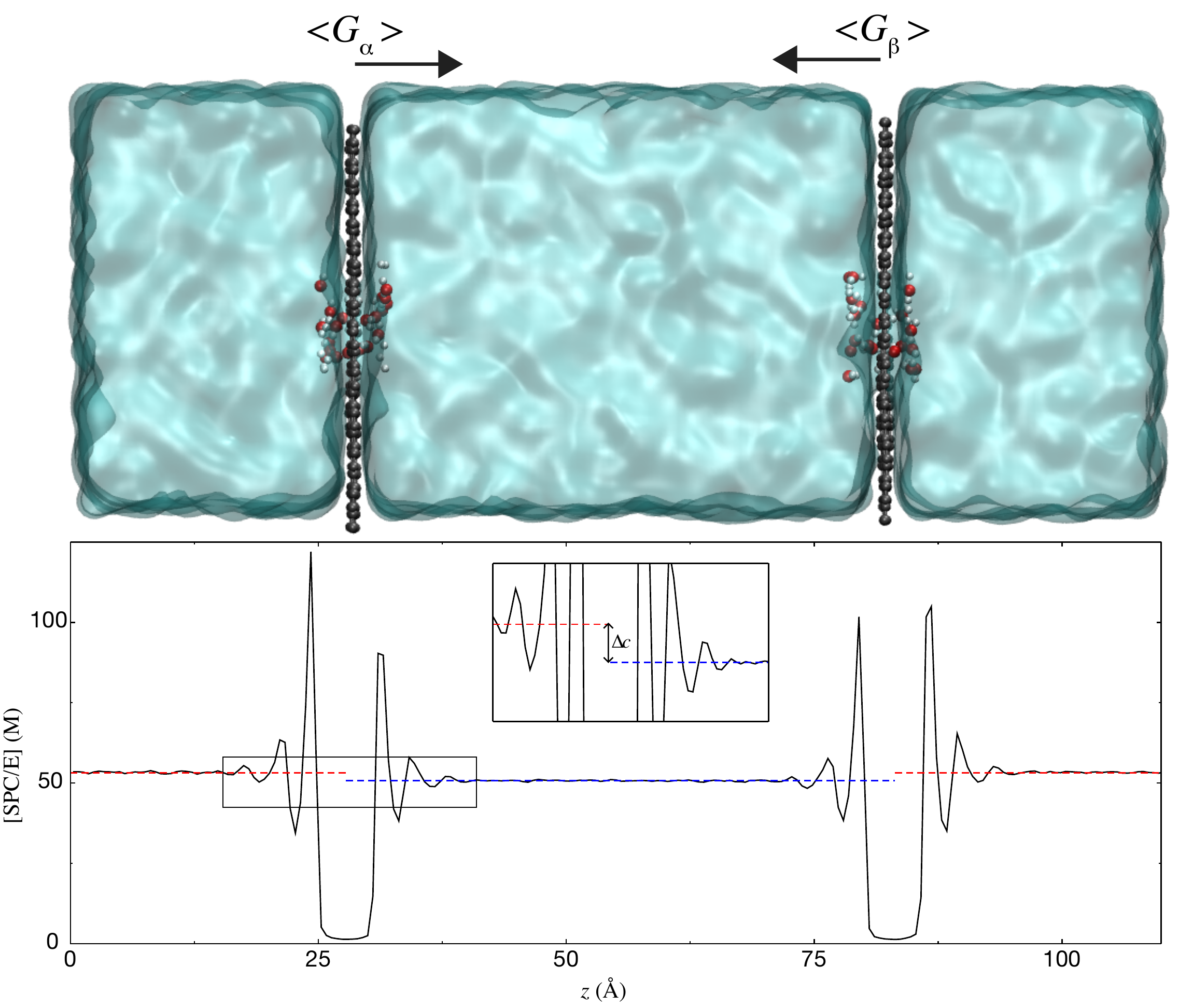}
    \caption{\label{fig:membrane} Top: A water box with two regions separated by nanoporous graphene sheets. Molecules close to the pores are indicated with a ball and stick representation. Under an applied molecular flux between the two regions, a steady state concentration difference develops between the two regions. Z-constraint forces are applied to counteract the net forces on the two membranes ($G_\alpha$ and $G_\beta$), and the arrows indicate time averages which are used to calculate the hydraulic pressure difference between the regions. 
    Bottom: the local concentration of water molecules in the box. Dashed horizontal lines indicate the mean water concentrations in the two regions. A magnified profile around one of the membrane regions highlights the concentration difference that develops in response to the molecular flux.}
\end{figure}

We have also tested the method on a molecular fluid with semipermeable membranes separating two regions of the $41 \angstrom \times 42 \angstrom \times 110 \angstrom$ simulation box. This system contains 5540 SPC/E water molecules (at 293.15 K and a density of $1 \mathrm{~g~cm^{-3}}$) which were modeled as 3-site rigid bodies.\cite{Berendsen:1987aa} The box was divided into two regions by two 41 \AA~$\times$ 42 \AA ~ graphene sheets that both contain hydrogen-terminated P28 pores, using the terminology adopted by Sun \textit{et al.}, where 28 refers to the number of disrupted graphene ring units by the addition of the pore. \cite{Sun2014} Parameters for the nanoporous graphene (NPG) carbon and hydrogen atoms were adopted from OPLS/2020.\cite{Jorgensen:2024aa} A full listing of all parameters used are given in the SI, and a representative snapshot of the system is shown at the top of Fig. \ref{fig:membrane}

In addition to the SPF-RNEMD molecular exchanges which migrate whole water molecules between the two regions, the two NPG sheets are held in place using Z-constraints, which fix the $z$ coordinates of the two NPG sheets with respect to the center of the mass of the system. This is a technique that was developed to obtain the forces required for the force auto-correlation calculations,\cite{Marrink94} originally proposed by Roux and Karplus to investigate the dynamics of ions inside ion channels.\cite{Roux91}
After the force calculation, the total force on molecule $\alpha$ is:
\begin{equation}
G_{\alpha} = \sum_i F_{\alpha i},
\label{eq:zc1}
\end{equation}
where $F_{\alpha i}$ is the force in the $z$ direction on atom $i$ in $z$-constrained molecule $\alpha$. The forces on the atoms in the $z$-constrained molecule are then adjusted to remove the total force on molecule $\alpha$:
\begin{equation}
F_{\alpha i} = F_{\alpha i} - 
	\frac{m_{\alpha i} G_{\alpha}}{\sum_i m_{\alpha i}}.
\end{equation}
Adjustments are also made to the velocities to keep the centers of mass of the NPG sheets fixed, and the accumulated constraint forces from both membranes are also subtracted from unconstrained liquid phase molecules to keep the system center of mass of the entire system from drifting. 

The average pressure on each of the two nanoporous graphene sheets can be computed easily from the time averages of the constraint forces that have been applied to keep them in place, so the mean pressure difference between the two regions,
\begin{equation}
\Delta P = P_\beta - P_\alpha = \frac{\left( \left< G_\beta \right> - \left< G_\alpha \right> \right)}{L_x L_y} .
\end{equation}
Note that $\Delta P$ is the hydraulic pressure difference that develops as a result of applying a steady-state particle flux between the two regions.

Once the system has come to steady state conditions, the flux of water across the membrane via diffusion is directly related to the applied particle flux,
\begin{equation}
    J_{wd} = J_p \times (2 L_x L_y) \times (v \rho)
\end{equation}
where $(2 L_x L_y)$ represents the dividing area between the two RNEMD regions, $v$ is the volume per water molecule, and $\rho$ is the pore density of the membrane.  The relevant linear constitutive equation for permeability, \cite{Merdaw2010, Cohen-Tanugi2014}
\begin{equation}
J_{wd} = A_{wd} \left(\Delta P - \Delta \Pi \right)
\label{eq:membraneFlux}
\end{equation}
involves both the hydraulic $(\Delta P)$ as well as the osmotic $(\Delta \Pi)$ pressure differences. Here, $A_{wd}$ is the pure water permeability by diffusion for this membrane.

Using the law of van't Hoff, the osmotic pressure difference can be calculated from the water concentration difference between the two regions, \cite{Lewis1908}
\begin{equation}
\Delta \Pi = RT \Delta c 
\end{equation}
which develops in response to the SPF-RNEMD exchanges. Therefore, from relatively straightforward SPF-RNEMD simulations, it is possible to apply a molecular flux between two regions, and from the concentration across the membrane and the time average of membrane constraint forces, obtain both the osmotic and hydraulic pressures.

\begin{figure}[H]
    \includegraphics[width=\linewidth]{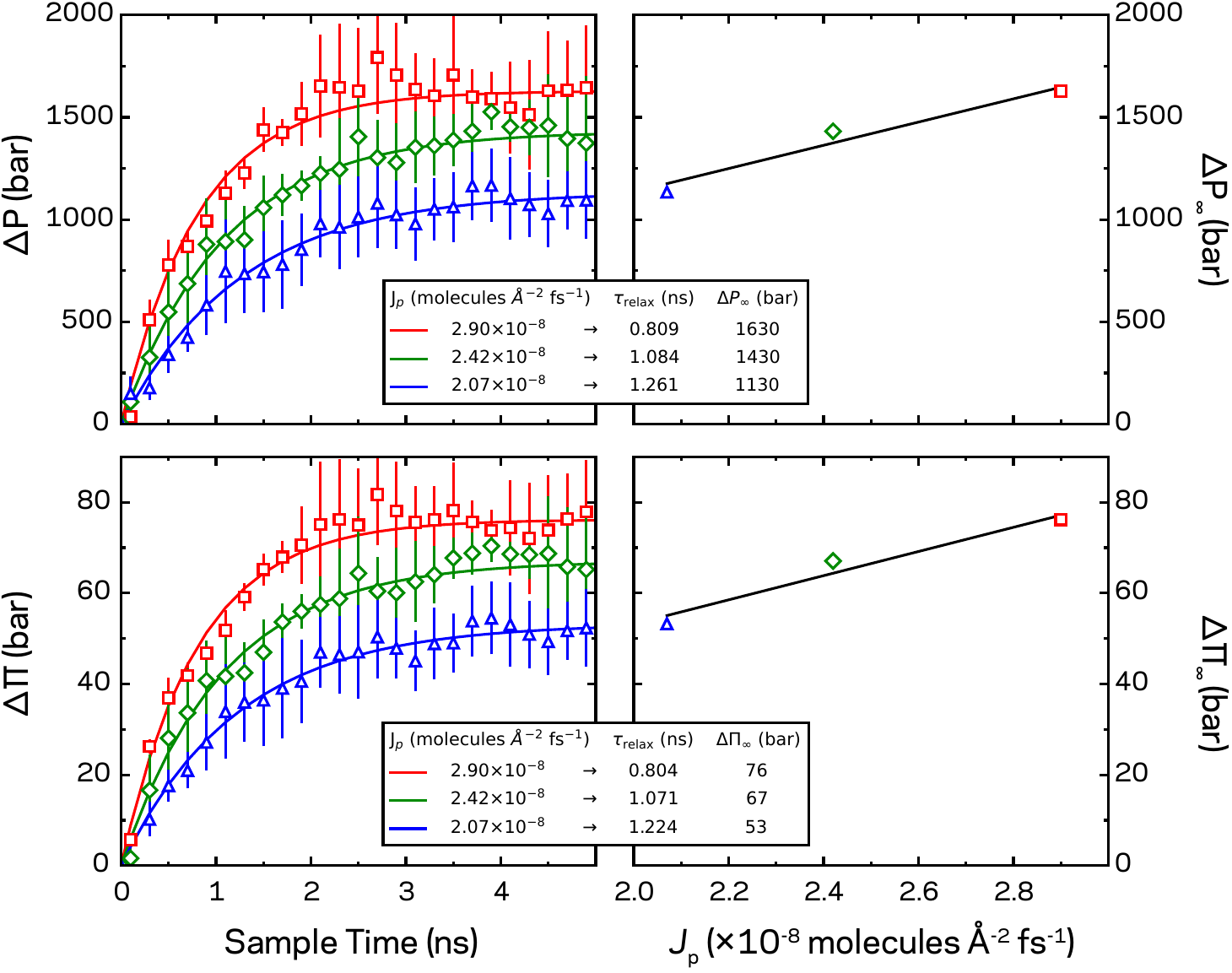}
    \caption{\label{fig:MembraneStitch} {For the nanoporous graphene / water system, convergence to steady-state hydraulic and osmotic pressure differences is reached after approximately 4 ns.  Here we have used applied flux values in the range  $2.07-2.90 \times 10^{-8}$ $\mathrm{molecules}~\angstrom^{-2} \mathrm{fs}^{-1}$. Data points are sampled from 200~ps windows, and 5 statistically independent replicas were started for each flux value. On the right side, we show the dependence of the infinite time $\Delta P$ and $\Delta \Pi$ values on the imposed flux.}} 
\end{figure}

In Fig. \ref{fig:MembraneStitch} we show the approach to the steady state $\Delta P$ and $\Delta \Pi$ for molecules in the two regions under various applied molecular flux values. Note that the hydraulic pressure is typically much larger than the osmotic pressure. Although we observe some local ordering of water near the NPG membrane walls, $\Delta c$ is calculated from the mean concentration of water in the two separated regions. Under very high fluxes (not included in Fig. \ref{fig:MembraneStitch}), the surface tension of water leads to bubble formation in the `source' region, but the flux values used here all yield stable concentration differences across the membrane. After the system has come to a steady state, the permeability is computed from a linear least squares fit on Eq. \eqref{eq:membraneFlux}.

The value of pure water permeability we have computed for the P28 pores, $A_{wd} = 6.85 \pm 0.15 \times 10^{-15}$ $\mathrm{L} \mathrm{~hr^{-1}} \mathrm{~bar^{-1}} \mathrm{~pore^{-1}}$.  Comparing with the calculations of \citeauthor{Cohen-Tanugi2014}, a pore density of $1.7 \times 10^{13} \mathrm{~cm^{-2}}$ would then yield a permeability of $1160 \pm 26 \mathrm{~L} \mathrm{~hr^{-1}} \mathrm{~m^{-2}} \mathrm{~bar^{-1}}$, which is reasonably close to the reported value of $1041 \pm 20 \mathrm{~L} \mathrm{~hr^{-1}} \mathrm{~m^{-2}} \mathrm{~bar^{-1}}$ for the larger pore in that study.\cite{Cohen-Tanugi2014}. 
\color{black}

\section{Conclusions}
We have presented a novel method for carrying out reverse non-equilibrium molecular dynamics simulations while simultaneously applying a particle flux or a thermal flux, or a combination of the two.  This method allows for straightforward calculation of isolated Fick diffusion constants in a mixture. The combined use of particle and thermal fluxes allows for calculation of the temperature dependence of diffusivity from a single simulation. Agreement with previous simulation methods is good. We note that one of our test cases, a mixture of identical, but distinguishable molecules provides an interesting result and test of the thermodynamic factor connecting Maxwell-Stefan and Fick diffusion constants. The differences in Kirkwood-Buff integrals for this system are subtle, but the effective entropy of mixing leads to some observable differences in the diffusivities of these systems that would not be apparent from self diffusion coefficients. 

\textcolor{black}{We note that for transport in uniform mixtures, the Maxwell-Stefan / Onsager approach using equilibrium molecular dynamics may be more useful for bulk diffusivities. The RNEMD method described above begins with the same Maxwell-Stefan definitions, but because a single particle flux is imposed independent of the other components, the Fick diffusivities arise naturally when applying the particle flux.}

\textcolor{black}{We have also tested the method on a system comprising a molecular fluid separated into two regions by a semipermeable membrane.  The method works well at computing concentration differences across the membrane, and coupled with Z-constraint forces to estimate hydraulic pressures, SPF-RNEMD allows for straightforward calculation of membrane permeability due to diffusion. Convergence of the hydraulic and osmotic pressures across the membrane was rapid compared with convergence to concentration gradients in bulk liquids.}

\textcolor{black}{In the semi-permeable membrane example, the ability to impose separate particle fluxes for a single component in a mixture is a feature of the method which will allow for calculation of single-species permeabilities, and potentially the osmotic pressure contribution due to solute concentration differences across the membrane. The timescales for convergence of the membrane permeability are significantly shorter than for bulk diffusivity, and using this method to study membrane separations of mixtures will be the subject of our future work.}

\begin{acknowledgement}
  Support for this project was provided by the National Science
  Foundation under grant CHE-1954648. Computational
  time was provided by the Center for Research Computing (CRC) at the University of Notre Dame.
\end{acknowledgement}

\begin{suppinfo}
\textcolor{black}{A derivation of the velocity scaling constraints (Eqs. \eqref{eq:a2} and \eqref{eq:b2}) from conservation laws is provided. We also include all other concentration profiles for the identical, but distinguishable, Lennard-Jones particles, as well as those for binary mixtures of Argon and Krypton. The SI also includes details on the simulations used to compute the Soret coefficient $(s_T)$ and force field parameters for nanoporous graphene sheets in SPC/E water.}
\end{suppinfo}

\newpage
\bibliography{spf}
\end{document}


\begin{abstract}
    This document contains a derivation of the velocity scaling coefficients (Eqs. (19) and (20) from the main text) using conservation of total energy.  We include all concentration profiles for the identical, but distinguishable, Lennard-Jones particles, as well as those for binary mixtures of Argon and Krypton. Also included are the details on the simulations used to compute the Soret coefficient $(s_T)$ utilized in the corresponding section of the manuscript. Lastly, tables with the force field parameters for nanoporous graphene sheets and SPC/E water are provided.
\end{abstract}

\pagebreak

\section{Derivation of the Velocity Scaling Constraints}

In Eqs. (19) and (20) in the main text, expressions for the velocity scaling parameters $a$ and $b$ are given. It may be helpful to the reader to understand the origin of these constraints.  Prior to a particle scaling move, the total potential depends on the instantaneous value of $\lambda$,
\begin{equation}
 U_\mathrm{before}(\mathbf{r}, \lambda) = \left[1-s(\lambda)\right] ~ U_\mathrm{source}(\mathbf{r}) + s(\lambda)~U_\mathrm{sink}(\mathbf{r}).\end{equation}
Similarly, before the scaling move, the kinetic energy in the two RNEMD exchange regions is given by,
\begin{align}
    K_{\textrm{before}} &= \frac{1}{2} \sum_{i = 1}^{N_a} m_i \textbf{v}_i^2 + \frac{1}{2} \sum_{j = 1}^{N_b} m_j \textbf{v}_j^2 \nonumber \\
    &= K_a + K_b \label{eq:K_before}
\end{align}
where $N_{\{a,b\}}$ are the numbers of atoms in regions a and b, $m_i$ is the mass and $\mathbf{v}_i$ is the velocity of atom $i$. After the particle scaling move (with the concomitant velocity scaling), the potential energy has been altered by incrementing $\lambda$,
\begin{equation}
 U_\mathrm{after}(\mathbf{r}, \lambda) = \left[1-s(\lambda + d\lambda)\right] ~ U_\mathrm{source}(\mathbf{r}) + s(\lambda + d\lambda)~U_\mathrm{sink}(\mathbf{r}).\end{equation}
and the kinetic energy has been altered by the two velocity scaling variables, $a$ and $b$,
\begin{align}
    K_{\textrm{after}} &= \frac{1}{2} \sum_{i = 1}^{N_a} m_i \left[ a \left( \textbf{v}_i - \langle \textbf{v}_a \rangle \right) + \langle \textbf{v}_a \rangle \right]^2 + \frac{1}{2} \sum_{j = 1}^{N_b} m_j \left[ b \left ( \textbf{v}_j - \langle \textbf{v}_b \rangle \right ) + \langle \textbf{v}_b \rangle \right]^2 \nonumber \\ 
    &= a^2 \left( K_a - \frac{1}{2} M_a \langle \textbf{v}_a \rangle^2\right) + \frac{1}{2} M_a \langle \textbf{v}_a \rangle^2 + b^2 \left( K_b - \frac{1}{2} M_b \langle \textbf{v}_b \rangle^2\right) + \frac{1}{2} M_b \langle \textbf{v}_b \rangle^2 ~ . \label{eq:K_after}
\end{align}
Here $M_a$ and $M_b$ are the total masses of particles in regions $a$ and $b$, respectively. Subtracting Eq. \eqref{eq:K_before} from Eq. \eqref{eq:K_after} gives the following:
\begin{equation}
    \Delta K = K_{\textrm{after}} - K_{\textrm{before}} = \left(a^2 - 1 \right) \left( K_a - \frac{1}{2} M_a \langle \textbf{v}_a \rangle^2\right) + \left(b^2 - 1 \right) \left( K_b - \frac{1}{2} M_b \langle \textbf{v}_b \rangle^2\right) ~ .
\end{equation}
It is possible to split the kinetic energy change into kinetic energy contributions from each of the RNEMD exchange regions,
\begin{align}
    \Delta K_a  &= \left(a^2 - 1 \right)\left( K_a - \frac{1}{2} M_a \langle \textbf{v}_a \rangle^2\right)  \\ \nonumber \\
    \Delta K_b &= \left(b^2 - 1 \right)\left( K_b - \frac{1}{2} M_b \langle \textbf{v}_b \rangle^2\right).
\end{align}
The difference in potential energy before and after the scaling move,
\begin{equation}
   \Delta U = U_\mathrm{after} - U_\mathrm{before} = \left[U_\mathrm{sink}(\mathbf{r}) - U_\mathrm{source}(\mathbf{r}) \right] \left[ s(\lambda + d\lambda) - s(\lambda) \right].
\end{equation}
In order to conserve total energy during the scaling moves, we require
\begin{equation}
    \Delta U + \Delta K = 0,
    \end{equation}
which implies
\begin{equation}
    \Delta K_a + \Delta K_b  = - \Delta U~.
\end{equation}
We also introduce an additional constraint equation which transfers a fixed amount of kinetic energy between the two regions. The amount of transferred kinetic energy is defined by the imposed heat flux $(J_q)$, the dividing area $(A)$ and the exchange time $(\Delta t)$,
\begin{equation}
    \Delta K_a - \Delta K_b = J_q~A~\Delta t
\end{equation}
These two constraint equations effectively set the allowed values of $a$ and $b$,
\begin{align}
2 \Delta K_a &= -\Delta U + J_q A \Delta t  = \left(a^2 - 1 \right)\left( 2 K_a - M_a \langle \textbf{v}_a \rangle^2\right)\\
2 \Delta K_b &= -\Delta U - J_q A \Delta  = \left(b^2 - 1 \right)\left( 2 K_b - M_b \langle \textbf{v}_b \rangle^2\right)
\end{align}
Solving Eqs. (12) and (13) for $a^2$ and $b^2$ then gives us
\begin{align}
    a^2 &= 1  - \frac{\Delta U - J_q~A~\Delta t}{2 K_a - M_a \left< \mathbf{v}_a \right>^2} \\ \nonumber \\
    b^2 &= 1  - \frac{\Delta U + J_q~A~\Delta t}{2 K_b - M_b \left< \mathbf{v}_b \right>^2} ~ .
\end{align}

\section{Identical (but distinguishable) molecules}
\begin{figure}[H]
    \includegraphics[width=\linewidth]{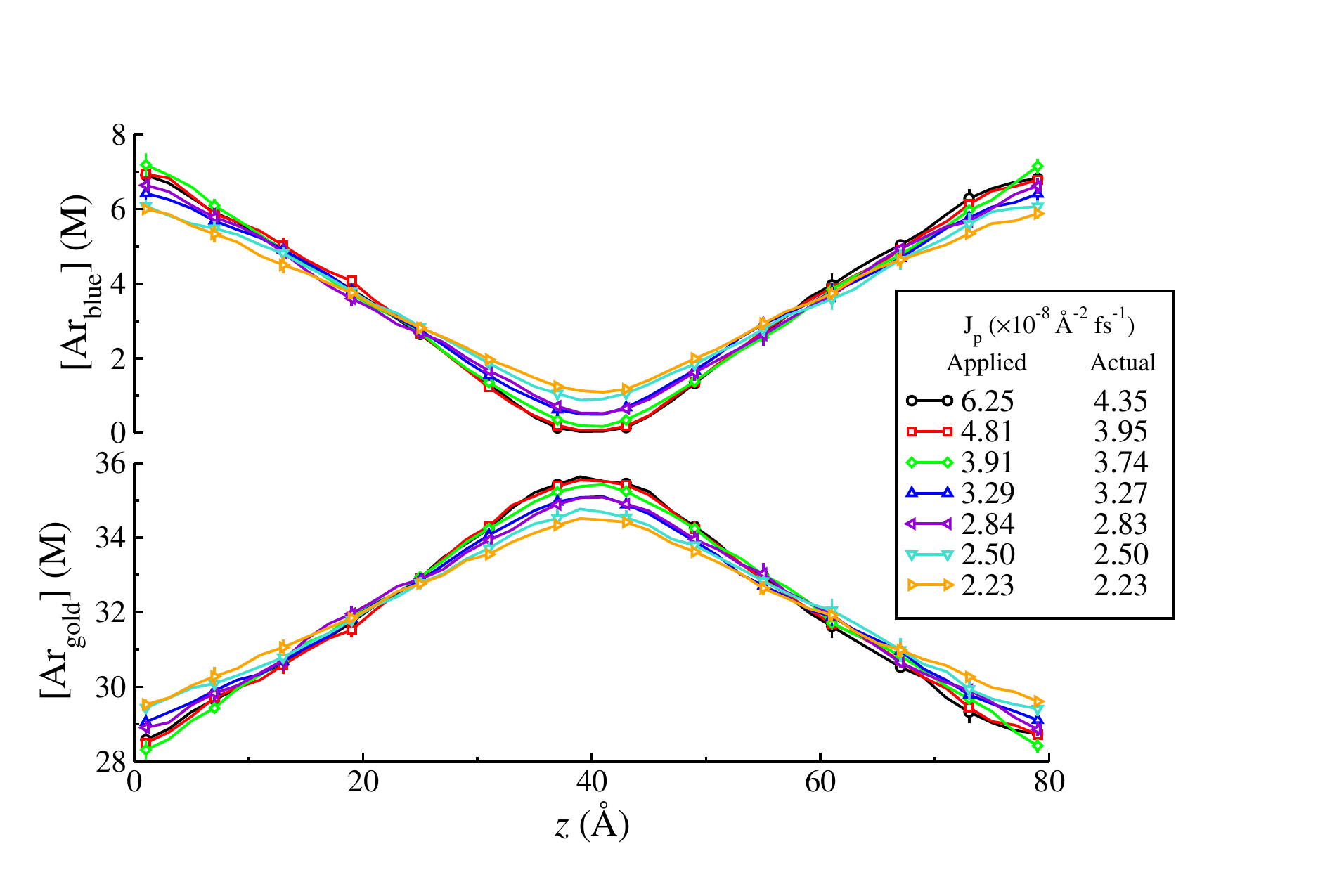}
    \caption{\label{fig:ArAr10_conc} Spatial concentration profiles for a 10:90 mixture of Ar$_\textrm{blue}$ and Ar$_\textrm{gold}$ under different applied particle fluxes ($J_p$), sampling from the last 2 ns of a 6 ns simulation. Error bars represent the 95\% confidence intervals for five statistically independent simulations. The difference between applied and actual particle flux is a result of unsuccessful RNEMD moves, caused by velocity scaling solutions which would result in changes larger than $0.1$\% of previous velocities.}
\end{figure}

\begin{figure}[H]
    \includegraphics[width=\linewidth]{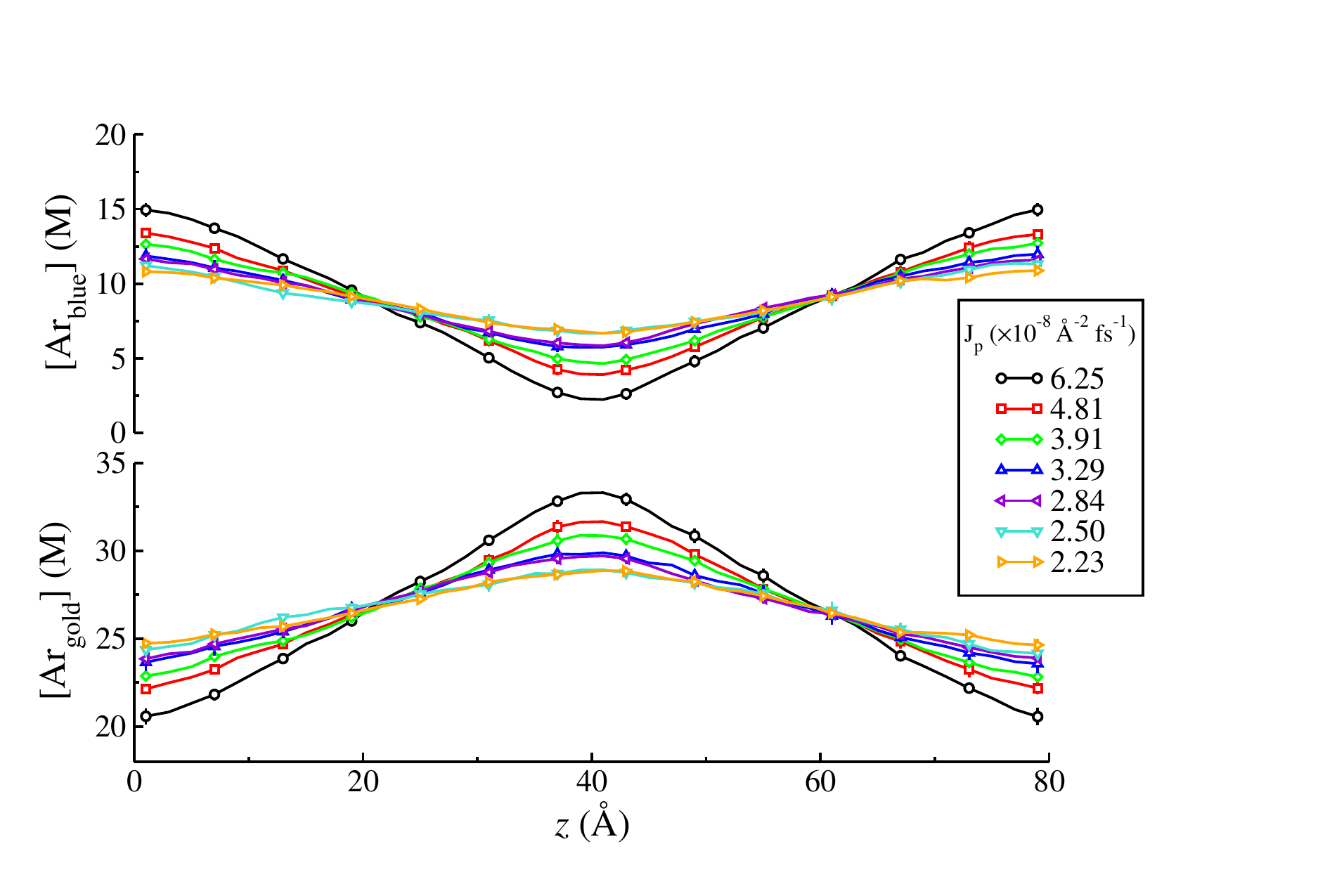}
    \caption{\label{fig:ArAr25_conc} Spatial concentration profiles for a 25:75 mixture of Ar$_\textrm{blue}$ and Ar$_\textrm{gold}$ under different applied particle fluxes ($J_p$), sampling from the last 2 ns of a 6 ns simulation. Error bars represent the 95\% confidence intervals for five statistically independent simulations.} 
\end{figure}

\begin{figure}[H]
    \includegraphics[width=\linewidth]{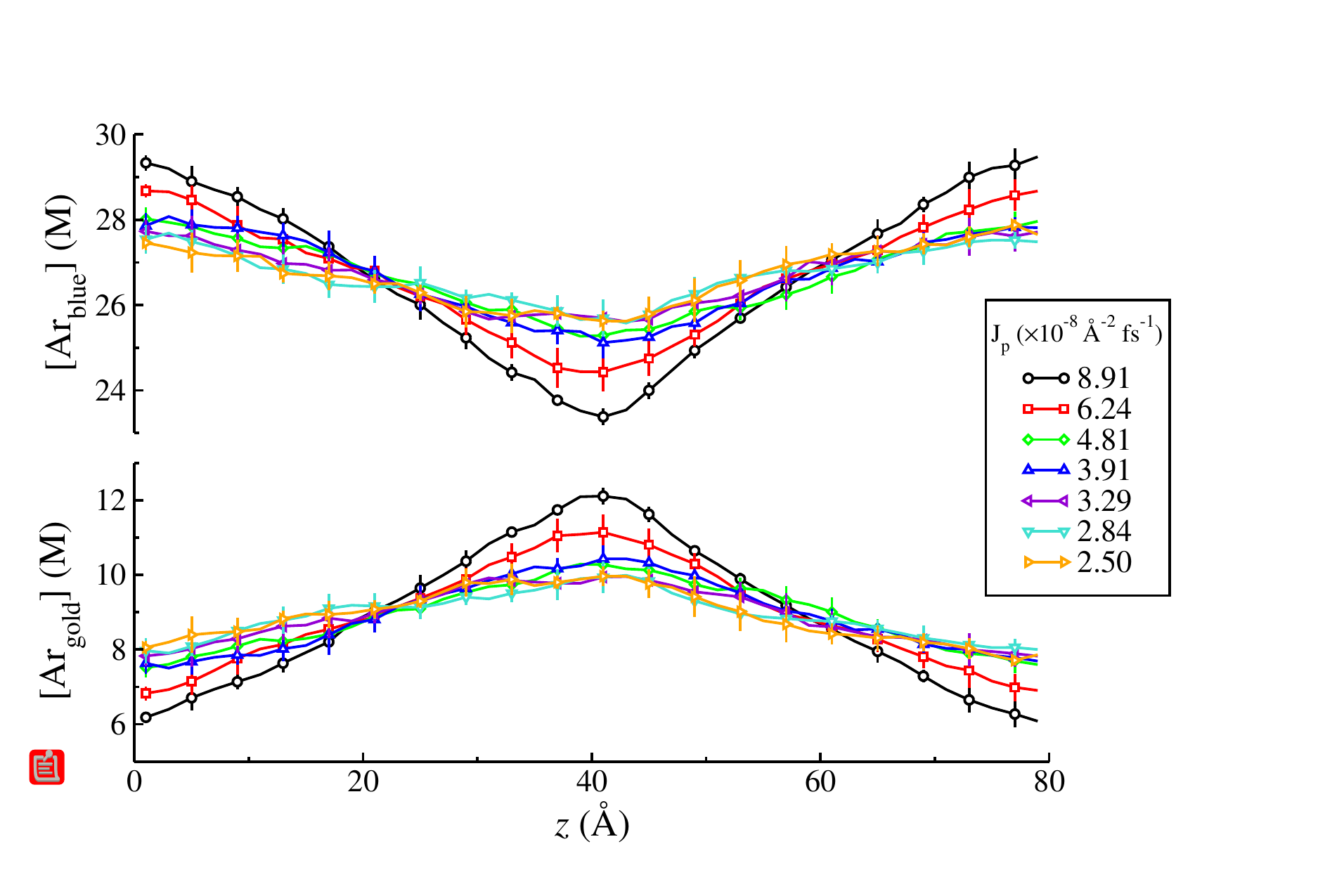}
    \caption{\label{fig:ArAr75_conc} Spatial concentration profiles for a 75:25 mixture of Ar$_\textrm{blue}$ and Ar$_\textrm{gold}$ under different applied particle fluxes ($J_p$), sampling from the last 2 ns of a 6 ns simulation. Error bars represent the 95\% confidence intervals for five statistically independent simulations.} 
\end{figure}

\begin{figure}[H]
    \includegraphics[width=\linewidth]{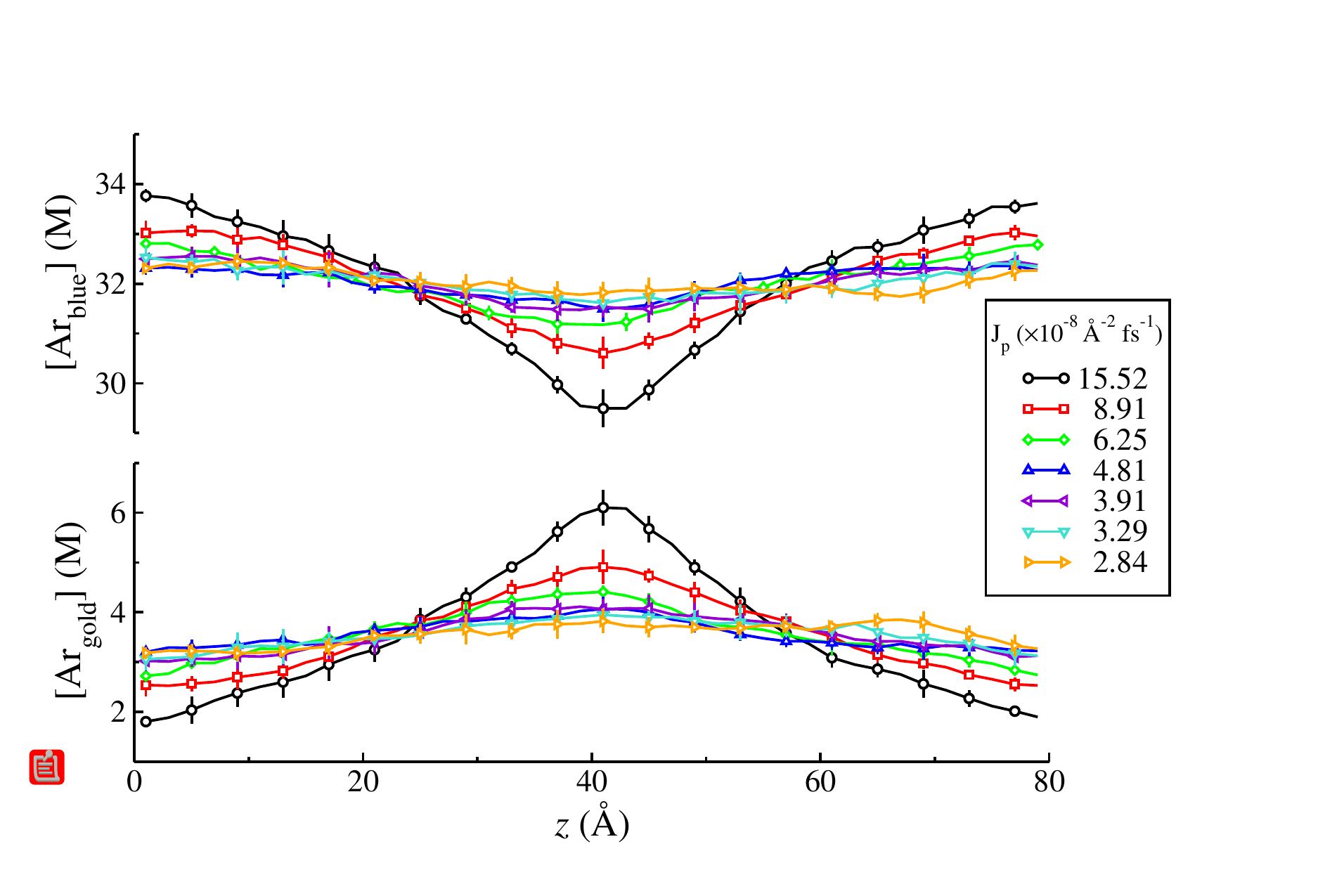}
    \caption{\label{fig:ArAr90_conc} Spatial concentration profiles for a 90:10 mixture of Ar$_\textrm{blue}$ and Ar$_\textrm{gold}$ under different applied particle fluxes ($J_p$), sampling from the last 2 ns of a 6 ns simulation. Error bars represent the 95\% confidence intervals for five statistically independent simulations.} 
\end{figure}

\section{Interdiffusion in a binary mixture of Lennard-Jones fluids}
  For systems that matched those studied by Yang \textit{et al.}, exchange periods are shown next to their corresponding particle flux. \cite{Yang2015} 
  
\begin{figure}[H]
    \includegraphics[width=\linewidth]{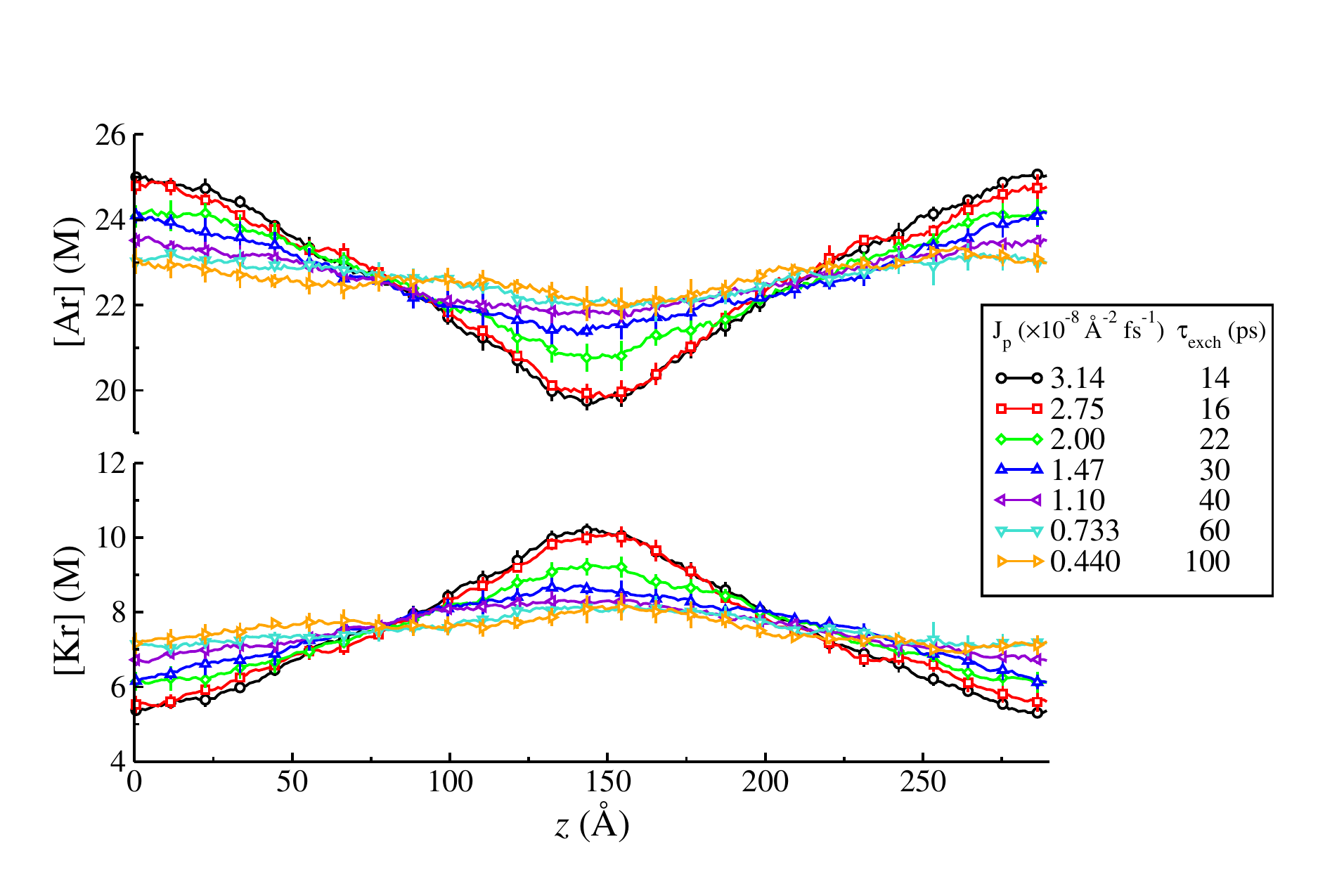}
    \caption{\label{fig:ArKr75_conc} Spatial concentration profiles for a 75:25 mixture of Ar and Kr under different applied particle fluxes ($J_p$), sampling from the last 10 ns of a 50 ns simulation. Error bars represent the 95\% confidence intervals for five statistically independent simulations. Equivalent particle exchange periods ($\tau_\mathrm{exch}$) are also shown for each value of the applied flux.} 
\end{figure}

\begin{figure}[H]
    \includegraphics[width=\linewidth]{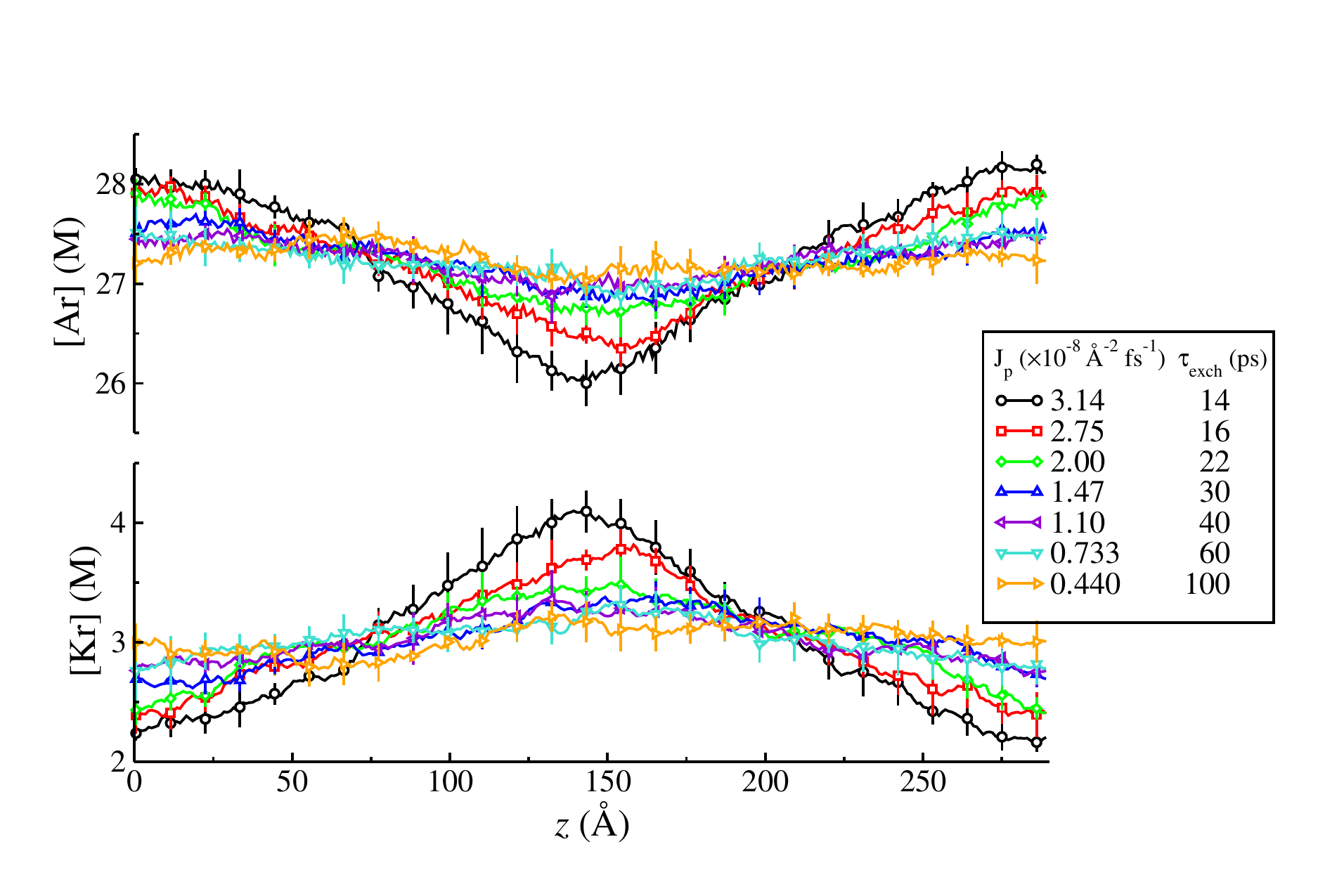}
    \caption{\label{fig:ArKr90_conc} Spatial concentration profiles for a 90:10 mixture of Ar and Kr under different applied particle fluxes ($J_p$), sampling from the last 10 ns of a 50 ns simulation. Error bars represent the 95\% confidence intervals for five statistically independent simulations. Equivalent particle exchange periods ($\tau_\mathrm{exch}$) are also shown for each value of the applied flux.} 
\end{figure}

\section{Temperature Dependence of Diffusion}
This section shows the result of applying a heat flux to a mixture and measuring the Soret coefficients from the resulting thermal and concentration gradients. 

\begin{figure}[H]    \includegraphics[width=\linewidth]{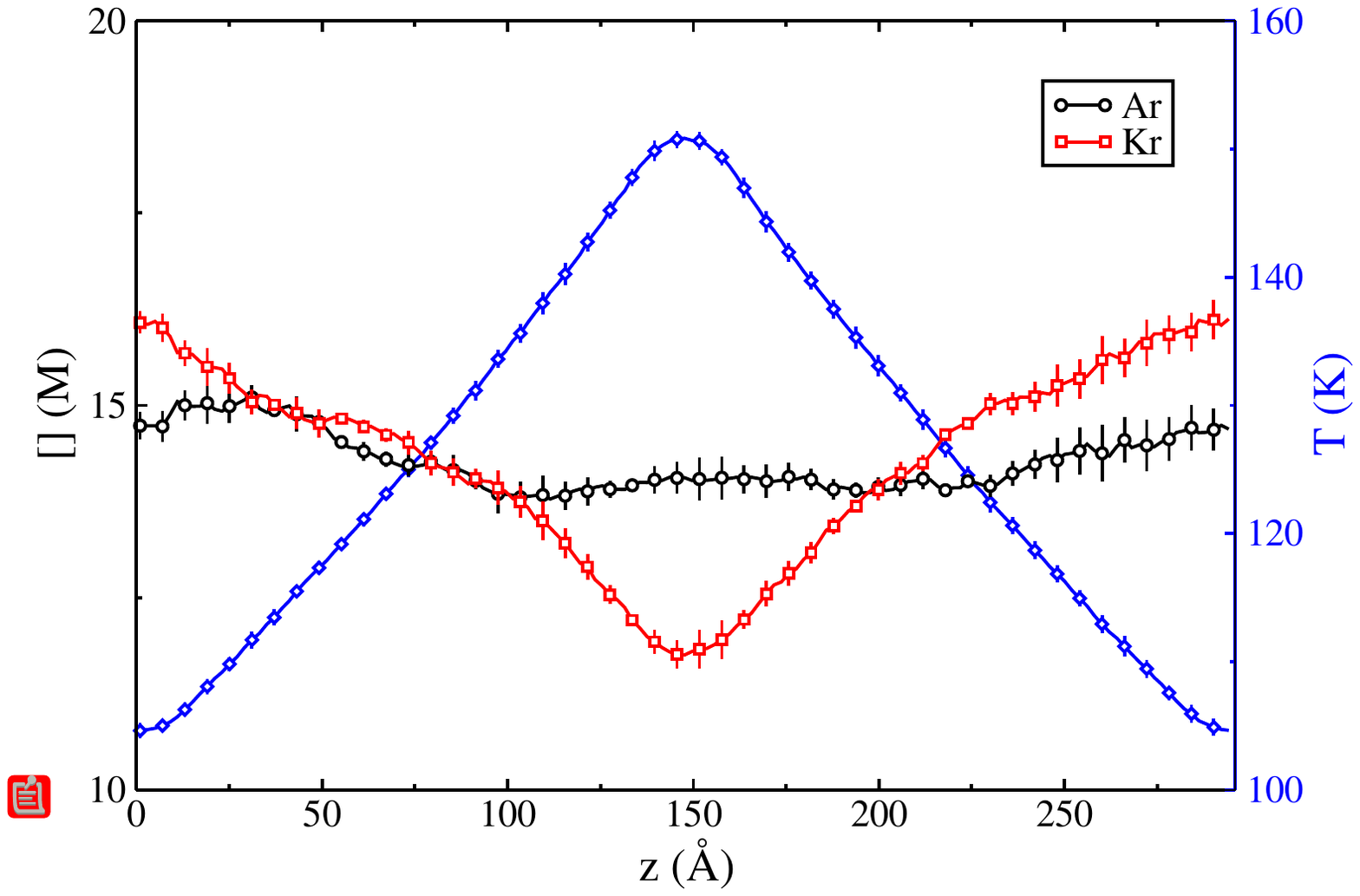}
    \caption{\label{fig:KineticFlux} Spatial concentration and temperature profiles for a 50:50 mixture of Ar and Kr operating under an applied thermal flux ($J_q$) of $4.0 \times 10^{-7}$ kcal mol$^{-1}$ \AA$^{-2}$ fs$^{-1}$ (278 MW m$^{-2}$). Error bars represent the 95\% confidence intervals for five statistically independent simulations. These simulations result in the measurement of two Soret coefficients: $3.65 \times 10^{-3}$ K$^{-1}$ and $11.4 \times 10^{-3}$ K$^{-1}$ for Argon and Krypton, respectively.} 
\end{figure}

As mentioned in the main text, when a mixture is evolving solely under a heat flux, it will reach steady state conditions and the concentration gradients will stabilize.  Starting with the coupled transport equations,
\begin{align}
\mathbf{J}_q &= -\lambda \nabla T -  k_B T^2 ~ \Gamma ~ D_T^D ~\nabla c_1 \\
\mathbf{J}_1 &= -c_t x_1 x_2 D_T^S \nabla T - D \nabla c_1
\end{align}
When the net flux in species 1 set to zero $(\mathbf{J}_1 = 0)$, simultaneous measurement of both concentration and thermal gradients therefore allows for the calculation of the Soret coefficient,
\begin{equation}
    s_T = \frac{D_T^S}{D} = - \frac{1}{c_t x_1 x_2} \left(\frac{\nabla c_1}{\nabla T}\right)~.
\end{equation}
In Fig. \ref{fig:KineticFlux} we show the thermal gradient superimposed with the two concentration gradients that arise when $J_q = 4.0 \times 10^{-7}$ kcal mol$^{-1}$ \AA$^{-2}$ fs$^{-1}$ (278 MW m$^{-2}$) is applied to the 50:50 Ar:Kr mixture.  Linear regression in the non-exchange regions provides estimates of both the temperature gradient $(\nabla T)$ and the concentration gradient $(\nabla c_\mathrm{Kr})$.  Our measured Krypton Soret coefficient agrees well with reported values, and due to the dominance of Krypton in the presence of Soret effects, Argon values are not usually reported.\cite{Reith2000, Perronace:2002ab}

\section{Molecular Separation Across Semi-permeable Membranes}

\subsection{Force Fields}
\begin{figure}[H]    \includegraphics[width=\linewidth]{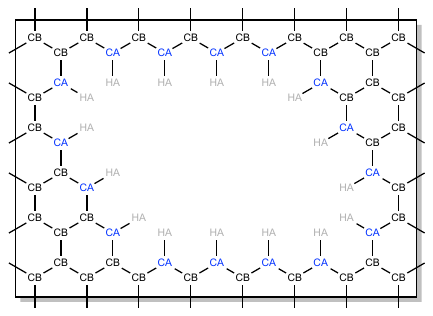}
    \caption{\label{fig:atomlabels} Atom types around the ``P28'' pore of a nanoporous graphene sheet. All other atoms in the graphene sheet are CB, and bonds are maintained across the periodic boundaries, so each sheet is a single molecule. The pore measures approximately $14.5 \angstrom \times 11.4 \angstrom$.} 
\end{figure}

The interaction parameters were adapted from multiple sources. 
The graphene parameters are from OPLS/2020.\cite{Jorgensen:2024aa} Rigid SPC/E water was used for the liquid phase.\cite{Berendsen:1987aa}  The atom types for nanoporous graphene are shown in Figure \ref{fig:atomlabels}, and the interaction parameters are shown in Tables \ref{tab:nonbonded} - \ref{tab:torsions}. 

In Table \ref{tab:bond}, harmonic bonds are described by
$V_\text{bond}(r) = \frac{k_{ij}}{2}\ (r_{ij}-r_{ij}^\circ)^2$
where $k_{ij}$ is the force constant and $r_{ij}^\circ$ is the equilibrium bond length. Water is simulated as a rigid body so no harmonic bond parameters are needed. The base atom types are shown here when applicable, so C refers to all carbon atoms that are not accounted for in the other listed harmonic bond parameters (CA and CB).

In Table \ref{tab:bend}, harmonic bends are in the form
$V_{\text{bend}}(\theta) = \frac{k_{ijk}}{2}(\theta - \theta_{ijk}^\circ)^2$ where $\theta$ describes the angle between adjacent atoms $i$, $j$, and $k$, where $j$ is the central atom.

In Table \ref{tab:torsions}, OPLS-style torsions are described by 
\begin{equation}
V_{\text{torsion}}(\phi) = \frac{1}{2} (v_1(1 + \cos\phi) + v_2(1 - \cos2\phi) + v_3(1 + \cos3\phi)) 
\end{equation}
although quartic torsions,
\begin{equation}
V_{\text{torsion}}(\phi) = k_0 + k_1\cos{\phi} + k_2\cos^2{\phi} + k_3\cos^3{\phi} + k_4\cos^4{\phi}.
\end{equation} 
are used internally in the OpenMD code for all torsion calculations.

For inversion potentials centered on $sp^2$-hybridized atoms with three satellites, the Amber improper torsion has the form 
\begin{equation}
V_\mathrm{improper}(\omega) = \frac{v}{2} \left( 1-\cos\left(2\left(\omega - \omega_0\right)\right) \right)~.
\end{equation}
where $\omega$ is an improper torsion angle described with the central atom in position 3 of a standard torsion, and the satellite atoms in positions 1, 2, and 4. Generally, $\omega_0$ is set to 0 for planar sites.  Here, we list the central atom of an inversion first.

\begin{table}[H]   
\bibpunct{}{}{,}{n}{,}{,}
\begin{center}
\caption{Non-bonded interactions used for simulations of nanoporous graphene sheets in SPC/E water} \label{tab:nonbonded}  
\begin{tabular}{ |c c c c c c| }      
\toprule                 
\text{Atom} & \text{Mass (au)} & $\sigma_{ii}$ (\AA{}) & $\epsilon_{ii}$ (kcal/mol) & Charge (e) & Source \\
\hline
\ce{\text{CA}} & 12.0107 & 3.55   & 0.068 & -0.115 & Ref. \protect\cite{Jorgensen:2024aa}\\
\ce{\text{CB}} & 12.0107 & 3.55   & 0.068 & 0 & Ref. \protect\cite{Jorgensen:2024aa}\\
\ce{\text{HA}} & 1.0079  & 2.42   & 0.030 & 0.115 & Ref. \protect\cite{Jorgensen:2024aa}\\
\ce{O-SPCE}    & 15.9994 & 3.16549 & 0.15532 & -0.8476 & Ref. \protect\cite{Berendsen:1987aa}\\
\ce{H-SPCE}    &  1.0079 & 0.0 & 0.0 & 0.4238 & Ref. \protect\cite{Berendsen:1987aa}\\
\hline
\end{tabular}
\end{center}
\end{table}
\begin{table}[H]                
\begin{center}
\bibpunct{}{}{,}{n}{,}{,}
\caption{Harmonic bond parameters for nanoporous graphene}
\label{tab:bond}
\begin{tabular}{ |c c c c c| }      
\hline                 
\emph{$i$} & \emph{$j$} & $r_{0}$ (\AA{}) & $k_{ij}$ (kcal mol$^{-1}$ \text{\AA{}}$^{-2}$) & Source\\
\hline
\ce{\text{CA}} & \text{HA} & 1.080 & 734.0 & Ref. \protect\cite{Jorgensen:2024aa}\\
\ce{\text{CB}} & \text{CB} & 1.370 & 1040.0 & Ref. \protect\cite{Jorgensen:2024aa}\\
\ce{\text{CA}} & \text{CB} & 1.404 & 938.0 & Ref. \protect\cite{Jorgensen:2024aa}\\
\ce{\text{CA}} & \text{CA} & 1.4   & 938.0 & Ref. \protect\cite{Jorgensen:2024aa}\\
\hline
\end{tabular}
\end{center}
\end{table}
\begin{table}[h]              
\begin{center}
\bibpunct{}{}{,}{n}{,}{,}
\caption{Harmonic bend parameters nanoporous graphene. The atom type C stands in for both CA and CB}
\label{tab:bend}    
\begin{tabular}{ |c c c c c c| }     
\hline                 
\emph{i} & \emph{j} & \emph{k} & $\theta_0$ (degrees) & $k_{ijk}$ (kcal mol$^{-1}$ \text{rad}$^{-2}$) & Source\\
\hline
\ce{\text{C}} & \text{C} & \text{C} & 120.0 & 126.0 & Ref. \protect\cite{Jorgensen:2024aa}\\
\ce{\text{C}} & \text{CA} & \text{HA} & 120.0 & 70.0 & Ref. \protect\cite{Jorgensen:2024aa}\\
\hline
\end{tabular}
\end{center}
\end{table}

\begin{table}[h]
\footnotesize
\begin{center}
\bibpunct{}{}{,}{n}{,}{,}
\centering
\caption{OPLS-style dihedral potentials for nanoporous graphene and Amber Improper torsion parameters for $sp^2$-hybridized carbon atoms}
\label{tab:torsions}    
\begin{tabular}{ cccc|ccccccl }
\toprule
$i$&$j$&$k$&$l$& Type & $v_1$ (kcal/mol) & $v_2$ (kcal/mol) & $v_3$ (kcal/mol) & Source\\
\midrule
\ce{\text{*}} & \text{C} & \text{C} &\text{*} & OPLS & 0.0 & 7.250 & 0.0 & Ref. \protect\cite{Jorgensen:2024aa}\\
\midrule
&&&&& $\omega_0 (^\circ)$ & $v$ (kcal/mol) & & \\
\midrule
\ce{\text{C}} & \text{*} & \text{*} &\text{*} & Amber Improper & 0.0 & {1.100} && Ref. \protect\cite{Jorgensen:2024aa}\\

 \bottomrule
\end{tabular}
\end{center}
\end{table}

\newpage

\bibliography{spf}